\documentclass[conference]{IEEEtran}
\IEEEoverridecommandlockouts

\usepackage{cite}
\usepackage{amsmath,amssymb,amsfonts}
\usepackage{algorithmic}
\usepackage{fontawesome5}
\usepackage{textcomp}
\usepackage{booktabs}
\usepackage{listings}
\usepackage{subcaption}
\usepackage{pifont}
\usepackage{tipx}
\usepackage{enumitem}
\usepackage{hyperref}
\usepackage{amsmath,amssymb,amsfonts}
\usepackage{algorithmic}
\usepackage{graphicx}
\usepackage{textcomp}
\usepackage{multirow}
\usepackage{makecell}
\usepackage{threeparttable}
\usepackage{siunitx}
\usepackage{tcolorbox}
\usepackage{xspace}

\def\BibTeX{{\rm B\kern-.05em{\sc i\kern-.025em b}\kern-.08em
    T\kern-.1667em\lower.7ex\hbox{E}\kern-.125emX}}

\newcommand{\xmin}{x_{min}}
\newcommand{\dstar}{\Delta^{*}}
\newcommand{\deltastep}{$\Delta$-stepping\xspace}
\newcommand{\dstarstep}{$\dstar$-stepping\xspace}
\newcommand{\rhostep}{$\rho$-stepping\xspace}

\tcbuselibrary{skins,xparse}

\definecolor{obsbg}{HTML}{EAF5EA} 
\definecolor{obsborder}{HTML}{00AA00} 

\newcounter{obscounter}

\newtcolorbox{observation}[1][]{ 
    colback=obsbg,
    colframe=obsborder,
    boxrule=0pt,
    leftrule=6pt,
    sharp corners,
    left=5pt, right=5pt, top=2pt, bottom=2pt,
    before skip=4pt,
    after skip=4pt,
    before upper={\refstepcounter{obscounter}\textbf{Observation \theobscounter:}~},
    #1 
}

\DeclareUnicodeCharacter{0394}{\ensuremath{\Delta}}
\DeclareUnicodeCharacter{2212}{\ensuremath{-}}

\begin{document}

\title{Mind the Gap: The Disconnect Between Synthetic and Natural Edge Weights in Parallel Single-Source Shortest Path%
\thanks{For the purpose of open access, the author has applied a Creative Commons Attribution (CC BY) license to any Author Accepted Manuscript version arising.}}

\author{%
\IEEEauthorblockN{\href{https://orcid.org/0009-0002-7399-7705}{Marco D'Antonio}}
\IEEEauthorblockA{Queen's University Belfast\\
Belfast, United Kingdom}
\and
\IEEEauthorblockN{\href{https://orcid.org/0000-0003-4599-1525}{Thai Son Mai}}
\IEEEauthorblockA{Queen's University Belfast\\
Belfast, United Kingdom}
\and
\IEEEauthorblockN{\href{https://orcid.org/0000-0001-5868-9259}{Hans Vandierendonck}}
\IEEEauthorblockA{Queen's University Belfast\\
Belfast, United Kingdom}}

\maketitle

\begin{abstract}
Scientific research works often evaluate Parallel Single-Source Shortest Path (SSSP) algorithms using synthetic, uniformly distributed edge weights.
However, real-world graphs exhibit very different, often heavy-tailed, weight distributions.
This creates a disconnect between how algorithms are evaluated and their real-world performance, since most SSSP implementations inherently rely on the weight distribution for parameter tuning and work efficiency.
In this paper, we explore whether current benchmarking methods unintentionally bias the performance results of these algorithms. 
To this end, we statistically characterize the weight distributions of 17 real-world graphs from a variety of domains and contrast them with six synthetic distributions used in the literature.
Through a comprehensive evaluation of seven state-of-the-art parallel SSSP algorithms, we demonstrate severe sensitivity to edge weights, and show that evaluating with synthetic uniform weights alters optimal parameter configurations and can invert the performance hierarchy.
These findings challenge existing benchmarking standards and offer practical insights for rigorous SSSP algorithm design.
\end{abstract}

\begin{IEEEkeywords}
Single-Source Shortest Path, Weighted Graphs, Benchmarking, Power-Law
\end{IEEEkeywords}

\section{Introduction}
Single-Source Shortest Path (SSSP) is a fundamental problem that has been thoroughly studied in the concurrent and parallel algorithms literature, producing implementations that are both theoretically efficient and fast in practice~\cite{bertsekasParallelAsynchronousLabelcorrecting1996, meyerDsteppingParallelizableShortest2003, shunLigraLightweightGraph2013, rihaniMultiQueuesSimpleRelaxed2015, dhulipalaJulienneFrameworkParallel2017, zhangOptimizingOrderedGraph2020, dhulipalaTheoreticallyEfficientParallel2021, dongEfficientSteppingAlgorithms2021, williamsEngineeringMultiQueuesFast2021, postnikovaMultiqueuesCanBe2022, zhangMultiBucketQueues2024, dantonioWaspEfficientAsynchronous2025}.
Driven by the proliferation of internet-scale applications and advancements in compute and storage infrastructures, graph datasets have grown exponentially, with trillion-scale graphs being utilized in private industry as early as 2015~\cite{chingOneTrillionEdges2015}.
However, publicly available, large-scale \textit{weighted} graph datasets suitable for benchmarking remain scarce.
For this reason, in order to assess the scaling of an algorithm on large datasets, the standard practice in the literature is to synthesize weights for naturally unweighted graphs.
Perhaps unsurprisingly, this has led to a fragmented landscape: a wide variety of synthetic weight distributions are employed across different studies, rarely with explicit justification regarding their representativeness of real-world workloads.
A prominent case is the Graph500 benchmark~\cite{murphy2010introducing}, which aims to ``provide useful information on the suitability of supercomputing systems for data intensive applications'', using two compute kernels as a benchmark: BFS and SSSP.
While the R-MAT~\cite{chakrabartiRMATRecursiveModel2004} topologies chosen for such benchmarks effectively mimic the scale-free behavior of real-world networks, their edge weights are assigned via a uniform distribution in $[0, 1)$.
As demonstrated in network science literature~\cite{yookWeightedEvolvingNetworks2001, barratArchitectureComplexWeighted2004, barratModelingEvolutionWeighted2004, liuQuantifyingEffectsTopology2018}, real-world edge weights rarely fit a uniform distribution; instead, they exhibit heavy-tailed behaviors.
This disconnect between network theory and HPC benchmarking practice raises a critical question: \textit{Are we doing SSSP benchmarking right?}

The discrepancy between natural and synthetic weights becomes even more pronounced when considering that most parallel SSSP algorithms are based on \deltastep~\cite{meyerDsteppingParallelizableShortest2003}, which groups vertices into buckets of width $\Delta$ based on their tentative distance from the source.
The $\Delta$ parameter regulates the trade-off between parallelism and redundant work, since the larger a bucket is, the more vertices can be processed in parallel.
However, the increased parallelism comes at a cost: vertices at greater distances may be processed concurrently with, or even before, those at shorter distances.
This inevitably leads to redundant work, as these vertices might be re-evaluated when shorter paths are subsequently discovered.
Consequently, fine-tuning $\Delta$ is critical for performance.
Because optimal bucket sizing depends on the distribution of path lengths, it is intrinsically linked to the underlying edge weight distribution of the graph.
This naturally raises the question of whether the underlying weight distribution fundamentally alters the performance profiles of various SSSP algorithms, potentially invalidating claims of highest practical performance derived from uniform distributions.
Furthermore, how do these distributions impact the tuning of $\Delta$ and similar algorithmic parameters? And crucially, do parameter-free SSSP algorithms demonstrate better robustness when using diverse weight distributions?

In this paper, we address these questions by establishing a ground truth based on real-world, naturally-weighted graphs.
We select 17 large-scale graphs, with up to 42 billion weighted edges, spanning road, traffic, social, semantic, and biological networks and characterize their empirical weight distributions.
Using established statistical methods~\cite{clausetPowerlawDistributionsEmpirical2009}, we fit theoretical distributions to both the body and the tail of the empirical data.
Against this baseline, we evaluate six synthetic weight distributions prevalent in the literature.
We then conduct a comprehensive performance analysis across seven state-of-the-art parallel SSSP implementations: five based on \deltastep\ (GBBS~\cite{dhulipalaTheoreticallyEfficientParallel2021}, GAP~\cite{beamerGAPBenchmarkSuite2017}, Wasp~\cite{dantonioWaspEfficientAsynchronous2025}, \dstarstep, and \rhostep~\cite{dongEfficientSteppingAlgorithms2021}), alongside two parameter-free approaches (MultiQueue-based parallel Dijkstra~\cite{rihaniMultiQueuesSimpleRelaxed2015} and parallel Bellman-Ford~\cite{bellmanRoutingProblem1958, mooreShortestPathMaze1959}).
We provide a comparison of the $\Delta$ tuning for each implementation and gather insights into how different implementations behave across varying topologies and weight distributions, highlighting large differences between natural and synthetic weights.
Furthermore, we compare the performance of fine-tuned implementations under different weight distributions, highlighting the large variety of results and how changing the weight distribution can fundamentally invert which algorithm achieves the highest practical performance.

In summary, the main contributions of this paper are as follows:
\begin{itemize}
    \item A statistical characterization of empirical weight distributions across a set of 17 naturally weighted graphs from a variety of domains.
    \item A comprehensive performance evaluation of seven state-of-the-art parallel SSSP implementations, analyzing how different graph topologies and weight distributions impact optimal $\Delta$ tuning.
    \item An empirical demonstration of the sensitivity of parallel SSSP algorithms to the edge weight distribution, alongside practical insights for users, performance analysts, and algorithm designers.
\end{itemize}

The rest of the paper is organized as follows.
Section~\ref{sec:background} presents the background on graphs, power-law distributions, and parallel SSSP, introducing the implementations compared in our study.
The experimental setup of our analysis is presented in Section~\ref{sec:setup}, along with the naturally-weighted graph datasets and the synthetic weight distributions used in literature.
Section~\ref{sec:characterization} characterizes the edge weight distribution of the naturally-weighted graphs datasets.
Section~\ref{sec:performance} analyzes the impact on performance of different weight distributions across the compared implementations.
The impact of weight distributions on parameter-based algorithms is presented in Section~\ref{sec:tuning}.
Section~\ref{sec:conclusion} concludes the paper with a discussion and insights for different SSSP algorithm practitioners.

\section{Background and Related Work}\label{sec:background}

\subsection{Graph Notation and Topologies}

Let $G$ be a directed weighted graph $G=(V,E,w)$ with a set of vertices $V$, a set of edges $E\subset V\times V$, and a weight function $w$.
Throughout this paper, $w$ can either be $w:E\to \mathbb{R}_{\geq0}$ or $w:E\to \mathbb{Z}_{\geq0}$, depending on the dataset or synthetic weight generation method.
Table~\ref{tab:weighted-datasets} provides a list of all the naturally-weighted graphs in this work, indicating whether they use integral or floating-point weights.
Where there is no ambiguity, we denote the number of vertices and edges as $n=|V|$ and $m=|E|$, respectively.

\subsection{Power-Law Distributions}
In network literature~\cite{barabasiEmergenceScalingRandom1999} a quantity $x$ obeys a power law if it is drawn from a probability distribution
\begin{equation*}
p(x)\sim x^{-\alpha},
\end{equation*}
where the $\sim$ symbol means ``roughly proportional''~\cite{voitalovScalefreeNetworksWell2019}.
Due to this ambiguous definition, empirical distributions are frequently mischaracterized as power laws, with classifications varying heavily depending on the chosen theoretical criteria or the fitting methodology~\cite{przuljModelingInteractomeScalefree2004, khaninHowScaleFreeAre2006, garciaSocialResilienceOnline2013, clausetPowerlawDistributionsEmpirical2009, broidoScalefreeNetworksAre2019, holmeRareEverywherePerspectives2019, voitalovScalefreeNetworksWell2019, articoHowRareAre2020}.
In this work, we use the definition of Clauset et al.~\cite{clausetPowerlawDistributionsEmpirical2009} in which a continuous power-law distribution is described by the following probability density:
\begin{equation*}
p(x)=\frac{\alpha-1}{\xmin}\left(\frac{x}{\xmin}\right)^{-\alpha},
\end{equation*}
where $\xmin$ is the lower bound of the power law, and $\alpha$ is a positive scaling parameter.
While this rigorous definition is often debated in the context of vertex degrees~\cite{voitalovScalefreeNetworksWell2019}, the statistical approach described by Clauset et al.~\cite{clausetPowerlawDistributionsEmpirical2009} is universally applicable to any empirical data.
An alternative definition for power-law distributions has been provided by Voitalov et al.~\cite{voitalovScalefreeNetworksWell2019}, encompassing all distributions whose probability density function is given by $p(x)=\ell(x)x^{-\alpha}$, where $\ell(x)$ is a slowly varying function, i.e. $\lim_{x\to\infty}\frac{\ell(tx)}{\ell(x)}=1$ for any $t > 0$.
In this context, the power law described by Clauset et al. is defined as a ``pure'' power law and represents the special case in which $\ell(x)$ is a constant.

\subsection{Parallel Single-Source Shortest Path}
We study edge weight distributions in the context of the  Single-Source Shortest Path (SSSP) problem, where given $G$ and a source vertex $s\in V$, the objective is to find the shortest distance from $s$ to any vertex $v$ that is reachable from $s$.
In a sequential setting, SSSP is traditionally solved by Dijkstra's algorithm~\cite{dijkstraNoteTwoProblems1959} which selects at each step the vertex $v$ with the shortest tentative distance $d[v]$ using a priority queue, and updates the distance of all of its neighbors, starting with $d[s]=0$.
Parallel SSSP algorithms generally fall into one of two execution paradigms: synchronous or asynchronous.

Synchronous approaches are predominantly based on the Bulk Synchronous Parallel (BSP) model~\cite{valiantBridgingModelParallel1990}, where computation proceeds in distinct supersteps separated by global synchronization barriers.
The most prominent algorithm in this category is $\Delta$-stepping~\cite{meyerDsteppingParallelizableShortest2003}, which groups vertices into buckets of width $\Delta$ depending on their tentative distance, i.e., bucket $i$ will hold vertices with distance in $[i\cdot \Delta, (i+1)\cdot\Delta)$.
A bucket is fully processed in parallel during a superstep before threads synchronize and move to the next.
The cost of parallelism, however, is redundant work, since vertices will not be processed following the work-efficient ordering of Dijkstra's algorithm.
Moreover, $\Delta$-stepping introduces an additional challenge: finding the optimal $\Delta$ value that balances parallelism against redundant work. The optimal $\Delta$ depends both on the graph topology and on the edge weight distribution, as we will point out later in the paper.
The parallel Bellman-Ford algorithm~\cite{bellmanRoutingProblem1958, mooreShortestPathMaze1959} also operates synchronously, offering a parameter-free alternative that converges also in the presence of negative weights, though it is generally slower in practice due to a higher volume of redundant vertex re-evaluations.

Alternatively, asynchronous models eliminate strict global barriers, allowing threads to process the active vertex frontier continuously and independently.
To achieve scalability without the strict ordering of $\Delta$-buckets, these implementations rely on highly concurrent data structures and execution strategies, such as relaxed priority queues~\cite{alistarhSprayListScalableRelaxed2015, rihaniMultiQueuesSimpleRelaxed2015, wimmerLockfreeKLSMRelaxed2015, postnikovaMultiqueuesCanBe2022} or work-stealing~\cite{nguyenLightweightInfrastructureGraph2013, postnikovaMultiqueuesCanBe2022, dantonioWaspEfficientAsynchronous2025}.
While asynchronous execution removes the overhead of global barriers, it allows threads to process vertices out of order. Because this can lead to a drastic increase in redundant work, many asynchronous approaches still rely on parameters like local $\Delta$ buckets to guide threads toward shorter paths.
Therefore, the efficiency of these algorithms is heavily dictated by their concurrent data structures, the tuning of their parameters, and how both react to the underlying edge weight distribution.

In this study, we analyze state-of-the-art parallel SSSP implementations. These include five algorithms based on $\Delta$-stepping (the GAP Benchmark Suite~\cite{beamerGAPBenchmarkSuite2017}, GBBS~\cite{dhulipalaTheoreticallyEfficientParallel2021}, \dstarstep, \rhostep~\cite{dongEfficientSteppingAlgorithms2021}, and Wasp~\cite{dantonioWaspEfficientAsynchronous2025}), one based on a relaxed priority queue (MultiQueue~\cite{rihaniMultiQueuesSimpleRelaxed2015, williamsEngineeringMultiQueuesFast2021}), and a parallel Bellman-Ford algorithm~\cite{bellmanRoutingProblem1958,dongEfficientSteppingAlgorithms2021}.
The implementations have several differences which in turn affect their performance on different kinds of graphs.
In the following we provide details about the implementations and their configuration used in our experiments.

\paragraph{\textbf{GAP}}
This framework implements a BSP-based \deltastep using thread-local buckets. At each superstep, the globally smallest bucket index is selected, and local vertices are moved to a global frontier to balance the workload.
GAP implements the bucket fusion optimization~\cite{zhangOptimizingOrderedGraph2020}, in which each thread processes the local content of the current bucket after processing the frontier in order to reduce synchronization costs.
We use the default threshold to continue bucket fusion if the local bucket has less than 1000 elements.

\paragraph{\textbf{GBBS}}
This implementation leverages the parallel bucketing structure introduced in Julienne~\cite{dhulipalaJulienneFrameworkParallel2017} and operates in a BSP fashion.
Its bucketing interface allows threads to extract all vertices mapped to a specific bucket in parallel and, following vertex label updates, to resize and update the buckets in parallel.
GBBS is configured with a fixed number of buckets.
We use the default value of 32, which performs better or the same as the value 128 used in the Julienne paper~\cite{dhulipalaJulienneFrameworkParallel2017}.

\paragraph{\textbf{\dstarstep\ and \rhostep}}
Both algorithms process all vertices up to a specific distance threshold during each step.
They differ primarily in how this threshold is calculated: \dstarstep increments the threshold by a fixed $\Delta$ at each step, whereas \rhostep dynamically sets the threshold to the $\rho$-th smallest distance within the current frontier using a sampling scheme.
The Lazy-Batched Priority Queue is introduced to support this framework and is implemented as a parallel hash-bag to extract and update vertices~\cite{wangParallelStrongConnectivity2023}.
The implementation has constants for the local queue size and the sample size in $\rho$-stepping; for such values we use the implementation defaults.

\paragraph{\textbf{Parallel Bellman-Ford}}
This is a BSP-based version of the classic Bellman-Ford algorithm. The implementation by Dong et al.~\cite{dongEfficientSteppingAlgorithms2021}
contains multiple optimizations: only processing active vertices, direction-optimization, and parallel sharing of large neighborhoods.
Since it does not use buckets, Bellman-Ford is not dependent on the $\Delta$ parameter.

\paragraph{\textbf{Wasp}}
The Wasp algorithm stores unresolved vertices across two data structures: a work-stealing-enabled active bucket and a thread-local list of future buckets.
Threads progress asynchronously: they first process their active bucket, then try to steal vertices with shorter tentative distances from other threads, and finally fall back to their thread-local lists if no higher-priority vertices are available.
Wasp buffers vertices in chunks and splits large neighborhoods in smaller chunks. 
The chunk size is the default of 64, and we also use the default threshold of $\theta=2^{20}$ for neighborhood splitting.

\paragraph{\textbf{The MultiQueue}}
This is a relaxed priority queue structure used to back an asynchronous parallel implementation of Dijkstra's algorithm.
It maintains $c \cdot  p$ lock-protected priority queues (8-ary heaps), where $p$ is the number of threads and $c$ is a scaling multiplier (we set $c=2$).
Vertices are extracted by randomly selecting two queues and extracting the higher-priority vertex, while generated vertices are pushed to a randomly chosen queue.
We use a high-performance implementation~\cite{williamsEngineeringMultiQueuesFast2021} in which threads stick to a certain queue for a number of trials, and use additional insertion and deletion buffers.
The size of the buffers is the default value of 16 elements.
Although stickiness is a parameter that significantly impacts performance~\cite{williamsEngineeringMultiQueuesFast2021}, we fix it at $s=64$ to evaluate the MultiQueue strictly as a parameter-free algorithm.

\section{Experimental Setup}\label{sec:setup}
In this section we present the naturally-weighted graphs datasets, the synthetic distributions gathered from the literature, and the setup of our experiments.

\subsection{Datasets}
Large-scale graphs are usually drawn from real-world networks such as road networks, social networks~\cite{newmanStructureScientificCollaboration2001,adamicSocialNetworkCaught2003a, backstromGroupFormationLarge2006, ahnAnalysisTopologicalCharacteristics2007, misloveMeasurementAnalysisOnline2007, chunComparisonOnlineSocial2008, leskovecPlanetaryscaleViewsLarge2008, edigerMassiveSocialNetwork2010a, kwakWhatTwitterSocial2010, garciaSocialResilienceOnline2013}, web crawls~\cite{albertDiameterWorldWideWeb1999, broderGraphStructureWeb2000, lehmbergGraphStructureWeb2014, meuselGraphStructureWeb2014, meuselGraphStructureWeb2015}, and biological networks~\cite{jeongLargescaleOrganizationMetabolic2000, barabasiNetworkBiologyUnderstanding2004, przuljModelingInteractomeScalefree2004, khaninHowScaleFreeAre2006, blumenthalEmergencePowerLaw2024}.
For our study, we selected real-world weighted graphs of sufficient size to benefit from parallel processing.
Table~\ref{tab:weighted-datasets} groups the datasets in road graphs and skewed-degree graphs.
Road graphs are typically characterized by a large diameter and a low, bounded average degree (e.g., the maximum out-degree in the \texttt{north-america} graph is 24, with an average degree of roughly 1).
Skewed-degree graphs are instead characterized by a small diameter and a larger average degree, and are often (loosely) categorized as scale-free networks, in which the degree distribution follows a power law.

\begin{table}
\centering
\caption{Naturally-weighted graph datasets used in the analysis. An overline on the graph abbreviation means the graph is undirected, an arrow means the graph is directed. $n$ is the number of vertices, $m$ is the number of directed edges -- every edge is counted twice in undirected graphs.}
\label{tab:weighted-datasets}
\begin{threeparttable}%
\begin{tabular}{llrr} 
\toprule
\textbf{Abbr.} &\textbf{Graph}  & \multicolumn{1}{c}{$\boldsymbol{n}$} & \multicolumn{1}{c}{$\boldsymbol{m}$} \\ \midrule
\multicolumn{4}{c}{\textbf{Road Graphs}} \\
$\overrightarrow{CA}$ & central-america   & 3 648 449      & 4 549 287      \\
$\overrightarrow{AO}$ & australia-oceania & 6 887 586      & 9 079 172      \\
$\overrightarrow{SA}$ & south-america     & 22 892 188     & 30 870 271     \\
$\overrightarrow{AF}$ & africa            & 35 279 191     & 47 019 425     \\
$\overrightarrow{NA}$ & north-america     & 95 654 077     & 122 973 777    \\
$\overrightarrow{AS}$ & asia              & 105 430 206    & 134 903 234    \\
$\overrightarrow{EU}$ & europe            & 141 140 217    & 180 706 815    \\
$\overrightarrow{PL}$ & planet            & 412 323 215    & 531 871 445    \\ 
\midrule
\multicolumn{4}{c}{\textbf{Skewed-degree Graphs}}    \\
$\overline{ARC}$ & archaea            &  1 644 227     & 204 792 654     \\
$\overline{EUK}$ & eukarya            &  3 243 106     & 359 763 936     \\
$\overline{IS}$ & isolates-sg1        &  34 982 171    & 20 981 181 510  \\
$\overline{MC}$ & metaclust           &  282 195 323   & 42 788 019 518  \\ 
$\overline{MW}$ & mawi$^\dagger$&  226 196 185   & 480 047 890     \\
$\overline{CA}$ & coauth-aminer$^\dagger$&  92 830 929    & 647 673 142     \\
$\overline{CM}$ & coauth-mag$^\dagger$&  173 195 937   & 1 089 154 246   \\
$\overline{CS}$ & colisten-spotify$^\dagger$&  3 604 455     & 1 927 482 013   \\
$\overline{ML}$ & moliere             &  30 239 687    & 6 677 301 366   \\ \bottomrule
\end{tabular}%
\begin{tablenotes}\footnotesize
  \item[$\dagger$] The graph has integer weights, in all other cases weights are single-precision floating point values.
\end{tablenotes}
\end{threeparttable}%
\vspace{-5mm}
\end{table}

\paragraph{\textbf{Road Graphs}} A prominent example of a weighted large-diameter graph is a road network, in which vertices represent intersections, edges represent streets, and weights represent street lengths.
One of the most frequently used graphs in the literature is the USA road network~\cite{beamerGAPBenchmarkSuite2017, dhulipalaJulienneFrameworkParallel2017, zhangOptimizingOrderedGraph2020, azadEvaluationGraphAnalytics2020} from the 9th DIMACS challenge~\cite{demetrescuShortestPathProblem2009a}; however, the dataset is from 2006 and the current North America dataset contains more than double the number of its edges.
For this reason, we converted the June 2025 versions of the OpenStreetMap (OSM) networks into the Matrix Market format by extracting the source and destination of each edge in the network along with its length.
The extracted node IDs were re-indexed from $1$ to $n$ in the Matrix Market format.
We release these datasets publicly with this paper\footnote{The datasets have been extracted using data from OpenStreetMap, available under the Open Database License.} 
These are, to the best of our knowledge, the latest converted versions of the OSM networks publicly available.

\paragraph{\textbf{Skewed-degree Graphs}}
While many skewed-degree graphs are used for evaluating performance in different benchmarks~\cite{leskovecRealisticMathematicallyTractable2005,murphy2010introducing, beamerGAPBenchmarkSuite2017}, most standard benchmark datasets are unweighted graphs.
The ten skewed-degree graphs listed in Table~\ref{tab:weighted-datasets} are sourced from a mix of general-purpose and specialized domain-specific repositories. 
The \texttt{mawi} and \texttt{moliere} datasets are obtained from the SuiteSparse Matrix Collection~\cite{davisUniversityFloridaSparse2011}, a standard repository for diverse graph and sparse matrix benchmarks.
The \texttt{mawi} dataset represents network traffic, where vertices are hosts and edge weights denote the number of packets exchanged between them. 
The \texttt{moliere} dataset models biomedical knowledge for hypothesis generation~\cite{sybrandtMOLIEREAutomaticBiomedical2017}.
Its vertices represent either a PubMed document, a Unified Medical Language System term, or an n-gram.
Edge weights represent semantic distance: the graph features explicit 0-weight edges to connect identical concepts, while larger values model weakly related concepts.
The \texttt{archaea}, \texttt{eukarya}, \texttt{isolates-sg1}, and \texttt{metaclust} datasets belong to the HipMCL data repository~\cite{azadHipMCLHighperformanceParallel2018}. 
All four are sequence similarity graphs, where vertices represent protein sequences and edge weights represent the similarity scores between them. 
The \texttt{coauth-aminer}, \texttt{coauth-mag}, and \texttt{colisten-spotify} datasets~\cite{tangArnetMinerExtractionMining2008, kumarRetrievingTopWeighted2020} originate from the Austin R. Benson dataset repository\footnote{Available online at \url{https://www.cs.cornell.edu/~arb/data/}.}. 
Among these, \texttt{coauth-aminer} and \texttt{coauth-mag} are co-authorship graphs, where vertices represent authors and edge weights denote the number of co-authored papers.
The \texttt{colisten-spotify} dataset was built starting from a large number of user streaming sessions, each with at most 20 songs.
In the weighted graph vertices represent songs, and the weight of an edge connecting two songs equals the number of sessions containing both songs.

\subsection{Synthetic Weight Distributions}

Table~\ref{tab:analyzed-dist} reports the different synthetic weight distributions we analyze for each graph, based on previous use in the literature.
The table highlights that recent benchmarks and studies have mostly relied on  uniformly distributed weights.
Specifically, the Graph500 benchmark~\cite{murphy2010introducing} assignes uniformly-distributed floating-point weights in $[0, 1)$, whereas other distributions proposed in other benchmarks~\cite{beamerGAPBenchmarkSuite2017} and shortest path literature~\cite{dhulipalaJulienneFrameworkParallel2017, dongParallelPointtoPointShortest2025} use integer weights.
A normal weight distribution has been previously used by Panitanarak et al.~\cite{panitanarakPerformanceAnalysisSinglesource2014}.
Because the paper does not explicitly state the parameters of their distribution, we assume a standard normal distribution.
Because our formulation of the SSSP problem requires non-negative edge weights, we take the absolute value of the generated weights, effectively creating a half-normal distribution.
Finally, we include a second normal distribution parameterized specifically for graph density, as suggested by an anonymous reviewer during a prior peer-review process.
The distribution has mean $\mu=1$ and standard deviation $\sigma=\sqrt{\frac{n}{m}}$; consequently, the standard deviation decreases as the graph density increases.
This distribution is also left-truncated to ensure weights are non-negative.
Additionally, a log-uniform distribution was used by Madduri et al.~\cite{madduriParallelShortestPath2006}, however,
we were unable to replicate their distribution as
the paper did not report the generation parameters.

\begin{table}
\centering
\caption{Edge weight distributions used in benchmarking.}
\label{tab:analyzed-dist}
\begin{threeparttable}
\begin{tabular}{lccc}
\toprule
\textbf{Type} & \textbf{Parameters} &  \textbf{Proposed in} & \textbf{Abbr.}\\ \midrule
\multirow{4}{*}{Uniform}& $a=0, b=1$\tnote{*} &  Murphy et al.~\cite{murphy2010introducing} & $U_{G500}$   \\
 & $a=1, b=255\tnote{$\dagger$}$ &Beamer et al.~\cite{beamerGAPBenchmarkSuite2017}  & $U_{BEAM}$             \\
 & $a=1,b=\lfloor log|V|\rfloor$\tnote{$\dagger$} &  Dhulipala et al.~\cite{dhulipalaJulienneFrameworkParallel2017} & $U_{DHUL}$\\
 & $a=1,b=2^{18}$\tnote{$\dagger$} & Dong et al.~\cite{dongParallelPointtoPointShortest2025} & $U_{DONG}$\\ \midrule
 \multirow{2}{*}{Normal}& $\mu=0$, $\sigma=1$& Panitanarak et al.~\cite{panitanarakPerformanceAnalysisSinglesource2014} & $N_{PANI}$\\
& $\mu=1$, $\sigma=\sqrt{\frac{n}{m}}$ & Anonymous Reviewers & $N_{ANON}$\\
\bottomrule
\end{tabular}
\begin{tablenotes}[para, flushleft]\footnotesize
  \item[*] The uniform distribution is defined in a right-open interval.
  \item[$\dagger$] The distribution has integer weights.
\end{tablenotes}
\end{threeparttable}
\vspace{-4mm}
\end{table}

\begin{figure*}[t]
    \centering
    \begin{subfigure}[b]{\textwidth}
        \centering
        \includegraphics[width=0.6\textwidth]{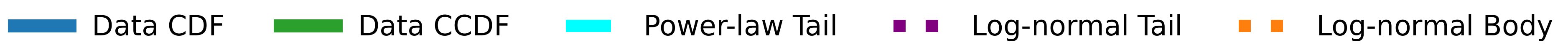}
    \end{subfigure}
    \begin{subfigure}[b]{0.23\textwidth}
        \centering
        \includegraphics[width=\textwidth]{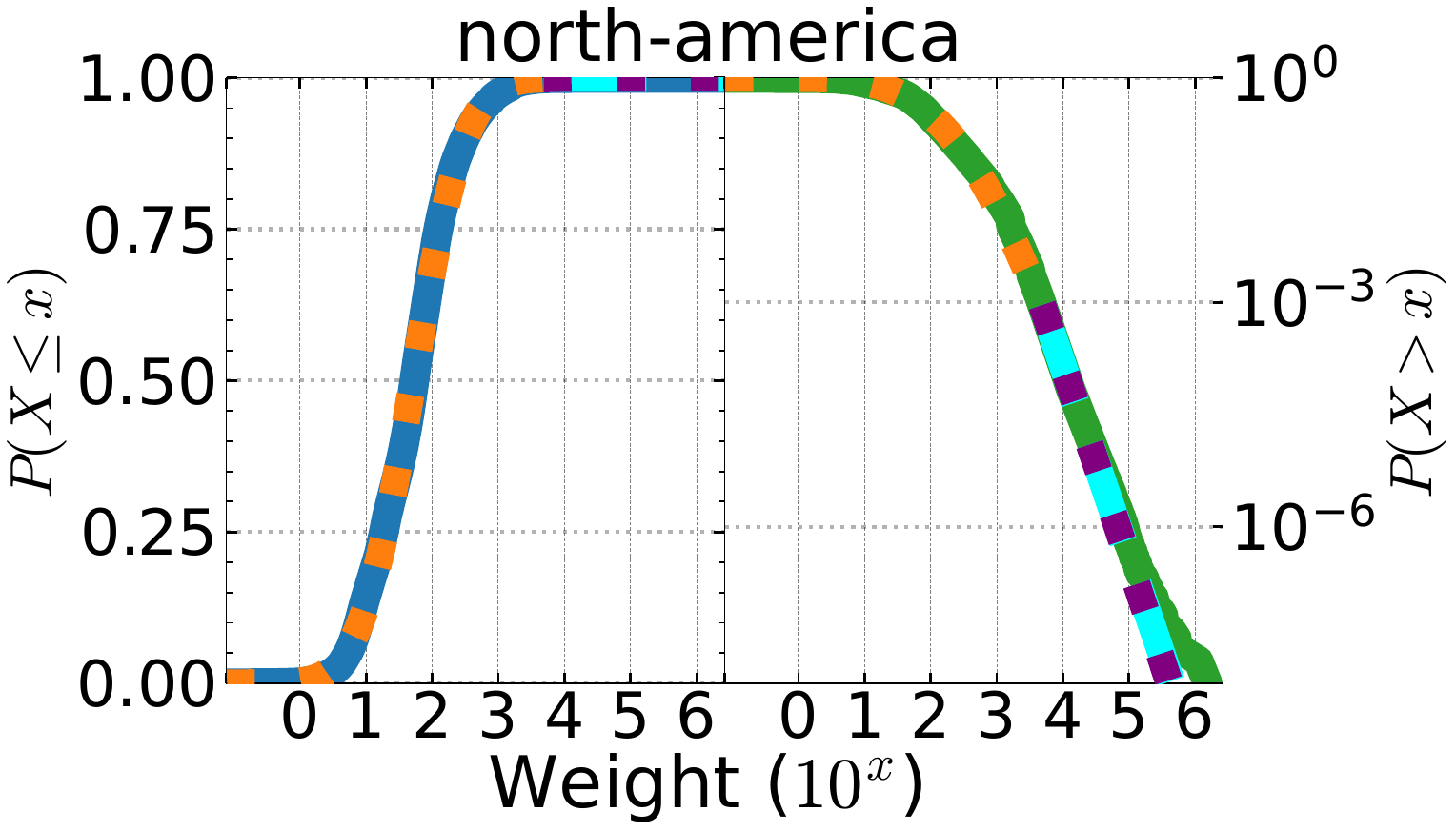}
    \end{subfigure}\hfill
    \begin{subfigure}[b]{0.23\textwidth}
        \centering
        \includegraphics[width=\textwidth]{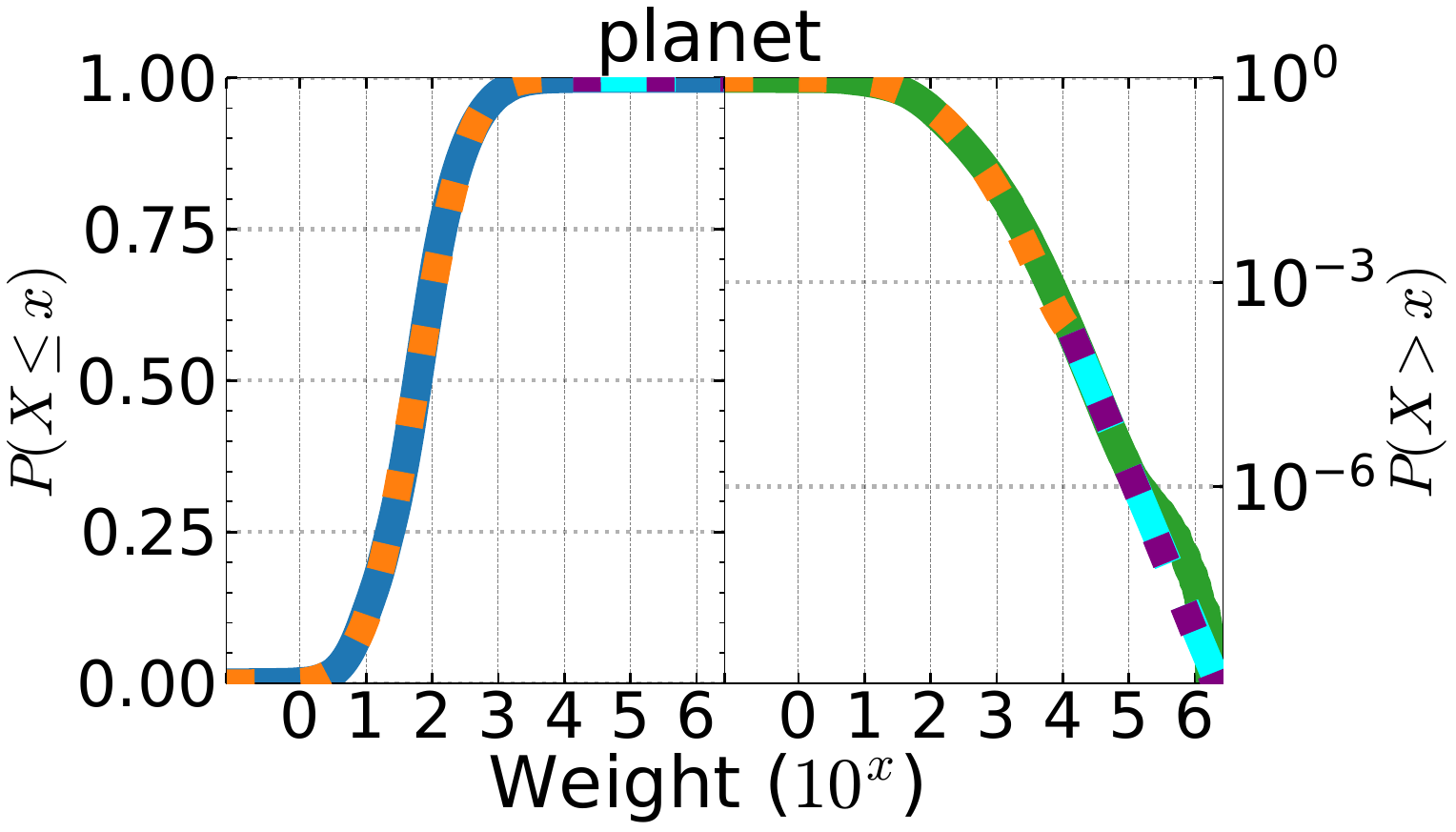}
    \end{subfigure}\hfill
    \begin{subfigure}[b]{0.23\textwidth}
        \centering
        \includegraphics[width=\textwidth]{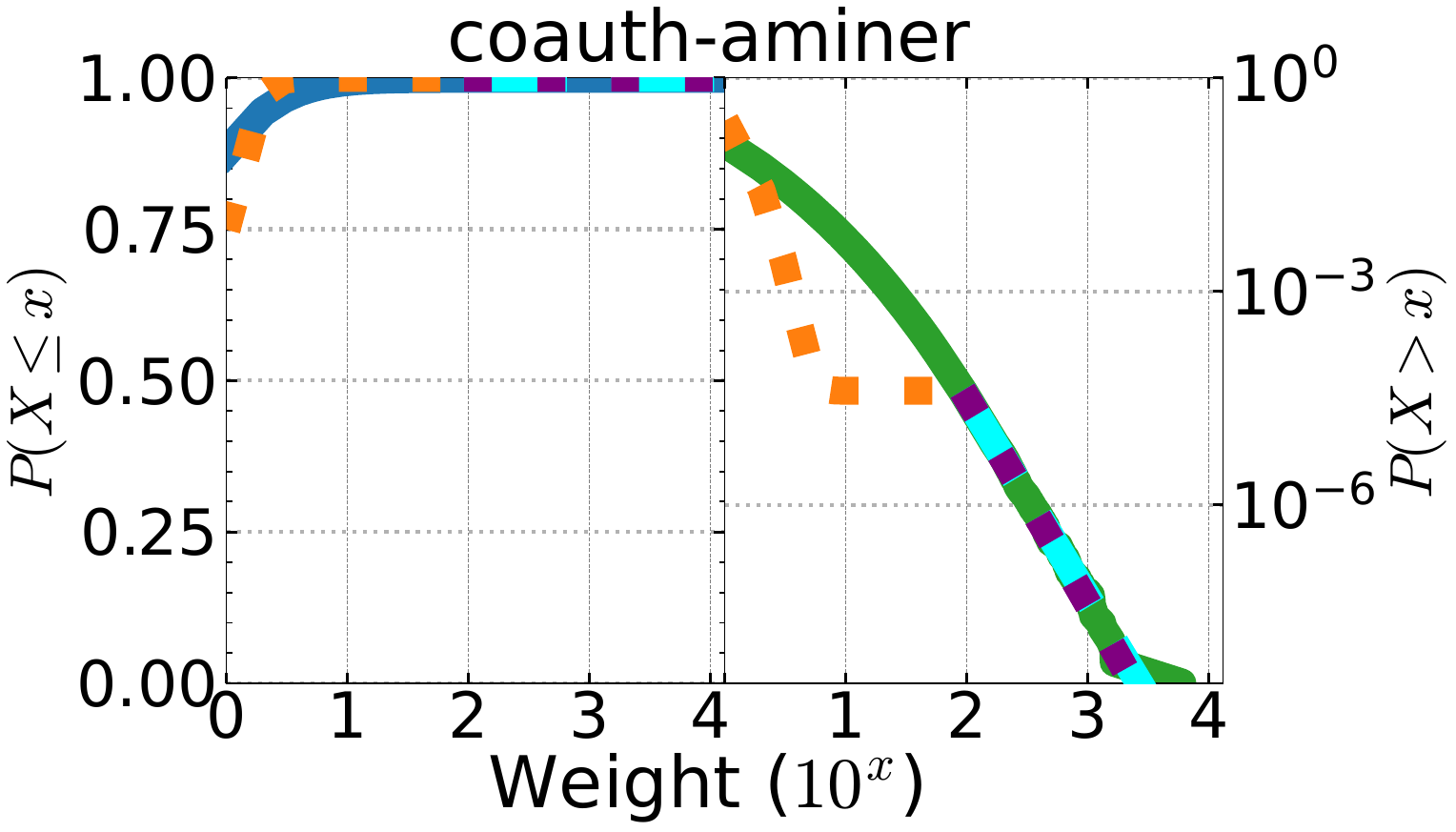}
    \end{subfigure}\hfill
    \begin{subfigure}[b]{0.23\textwidth}
        \centering
        \includegraphics[width=\textwidth]{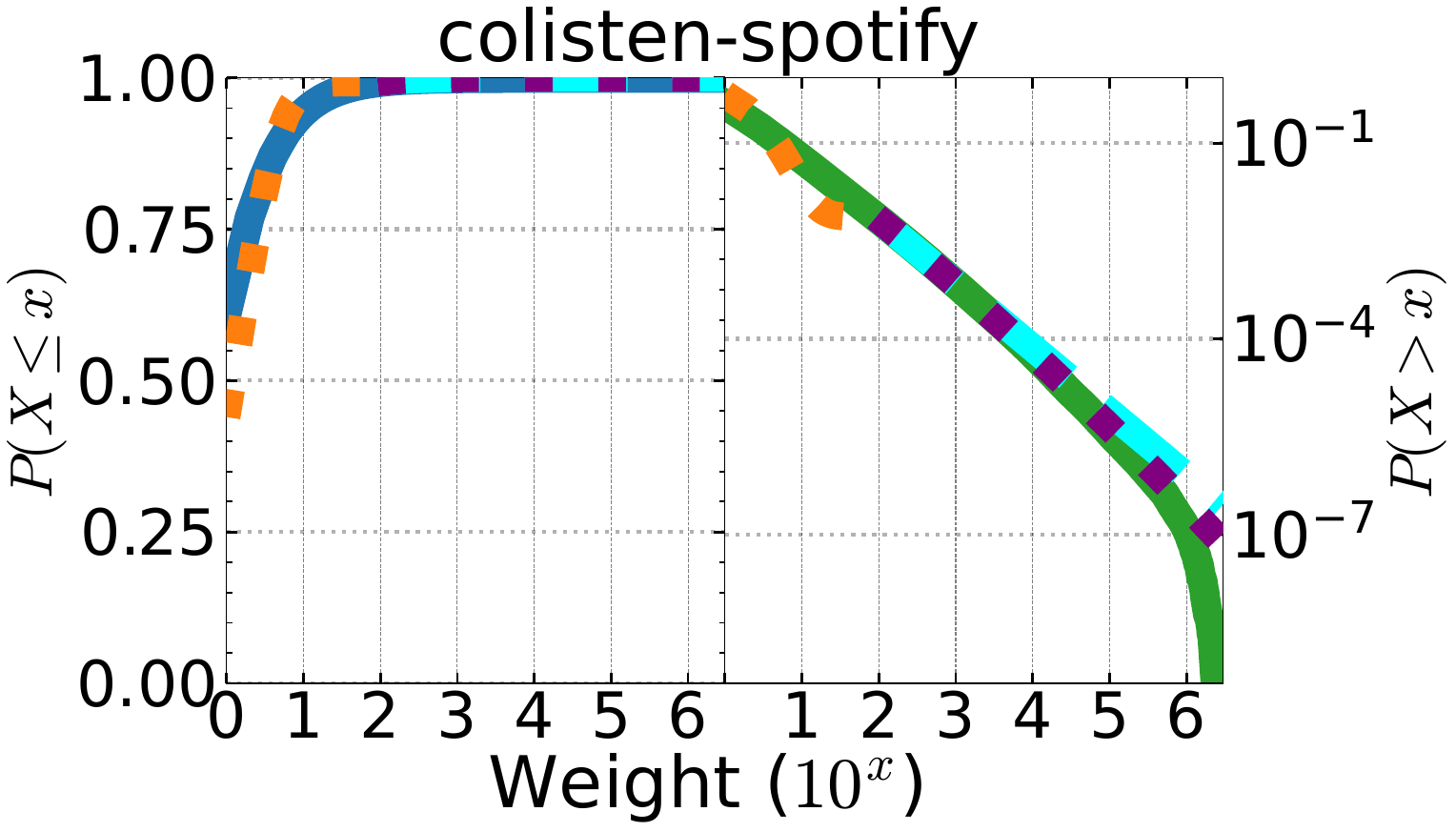}
    \end{subfigure}\hfill
    \begin{subfigure}[b]{0.23\textwidth}
        \centering
        \includegraphics[width=\textwidth]{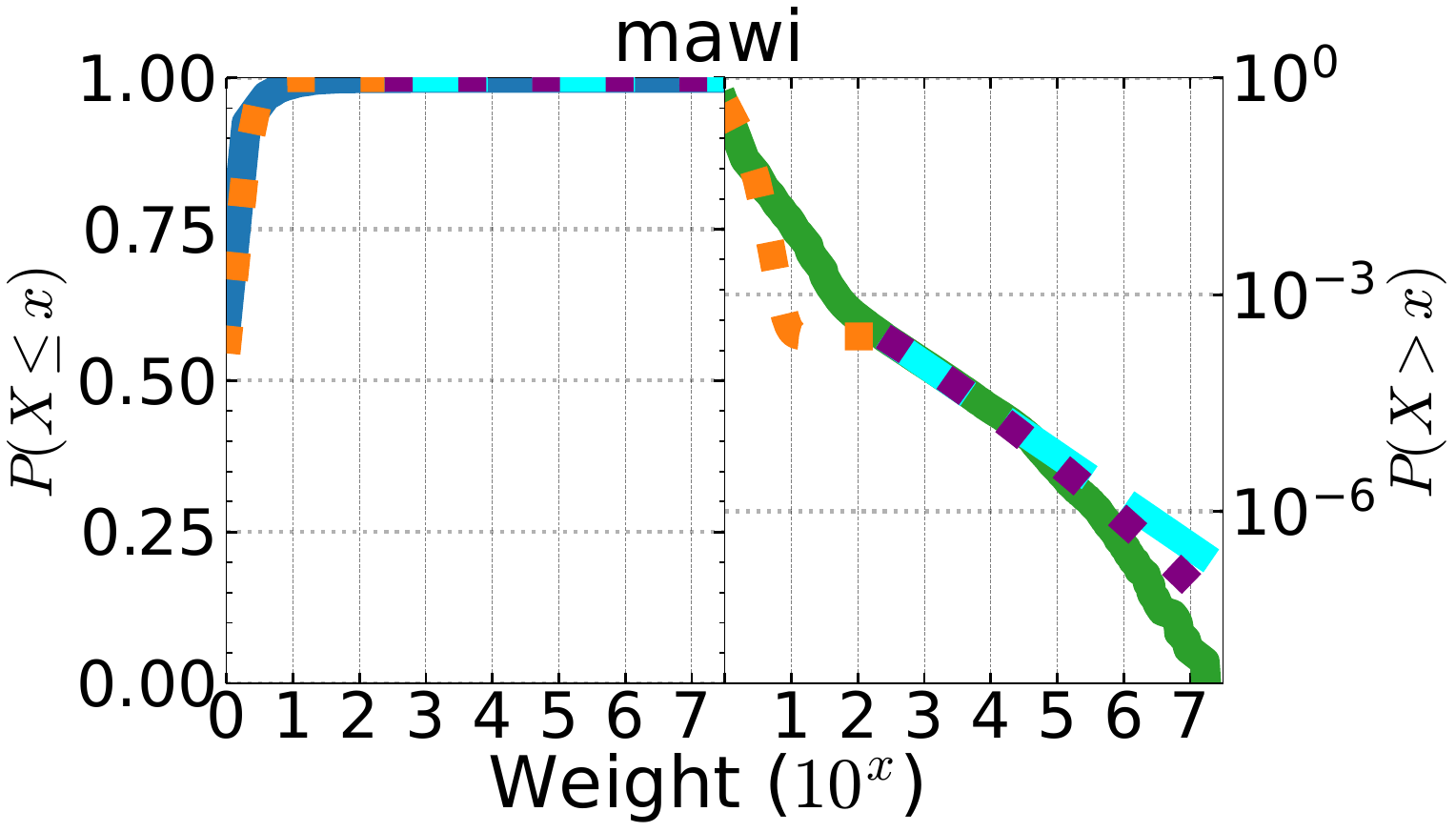}
    \end{subfigure}\hfill
    \begin{subfigure}[b]{0.23\textwidth}
        \centering
        \includegraphics[width=\textwidth]{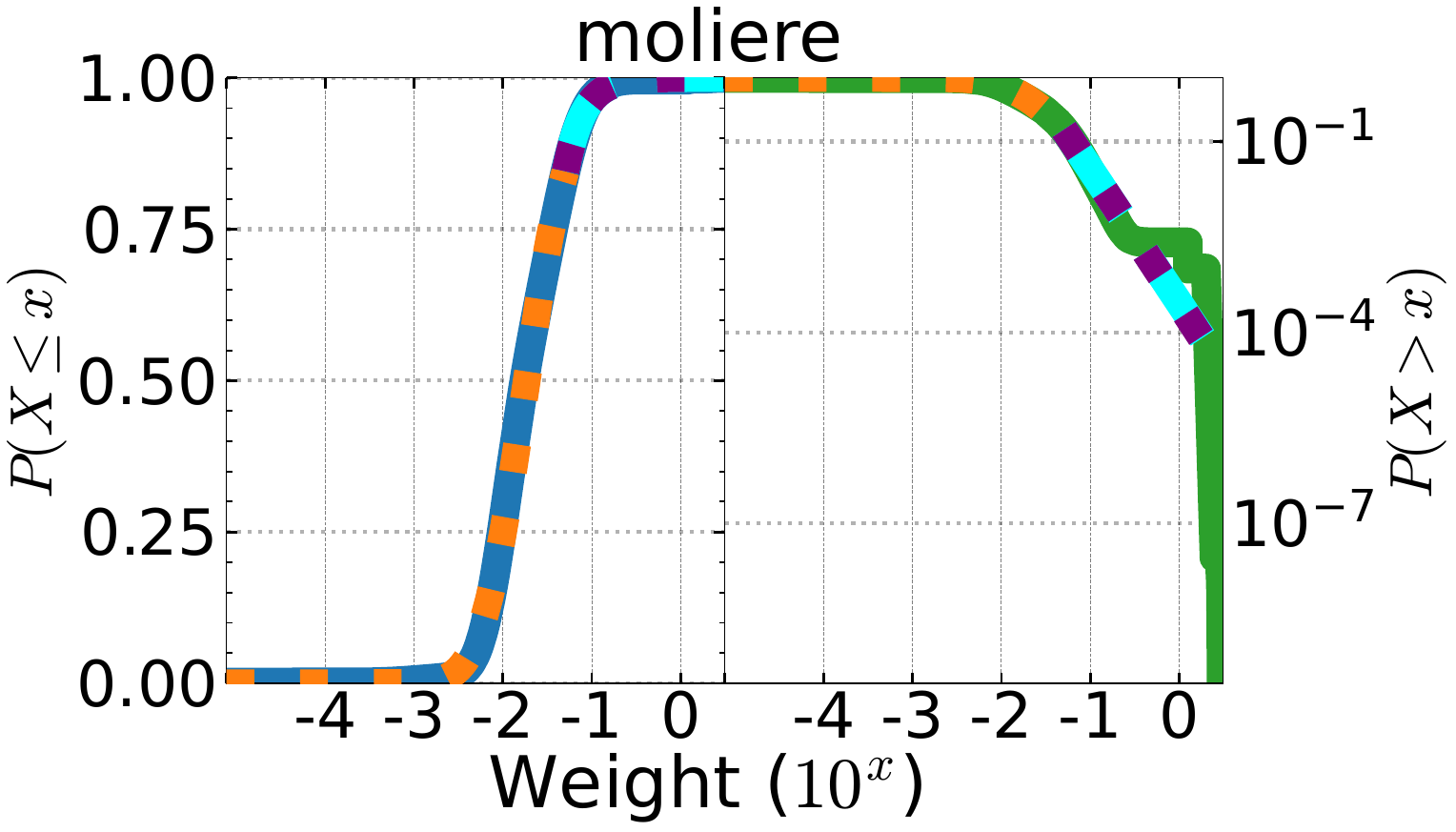}
    \end{subfigure}\hfill
    \begin{subfigure}[b]{0.23\textwidth}
        \centering
        \includegraphics[width=\textwidth]{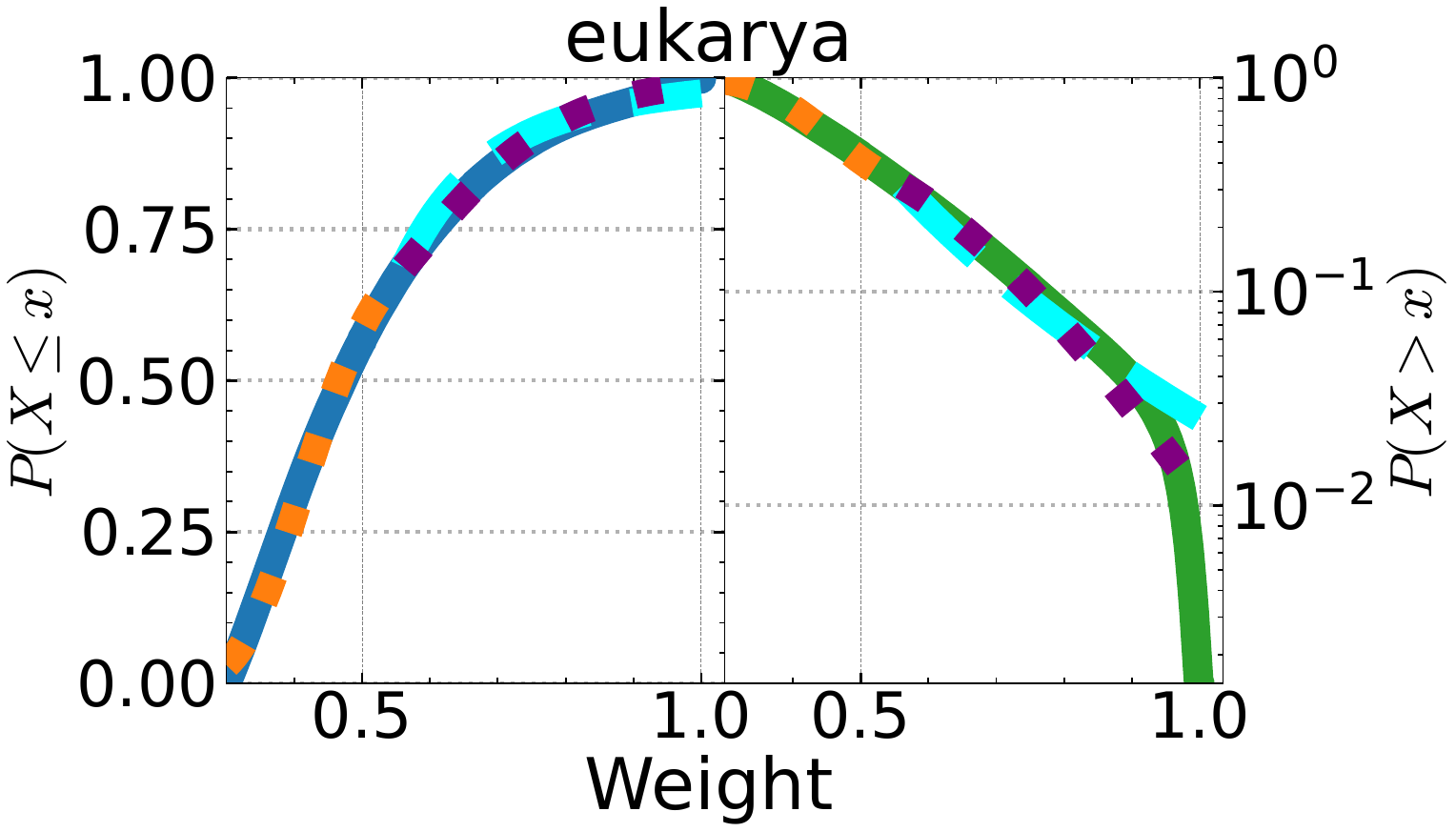}
    \end{subfigure}\hfill
    \begin{subfigure}[b]{0.23\textwidth}
        \centering
        \includegraphics[width=\textwidth]{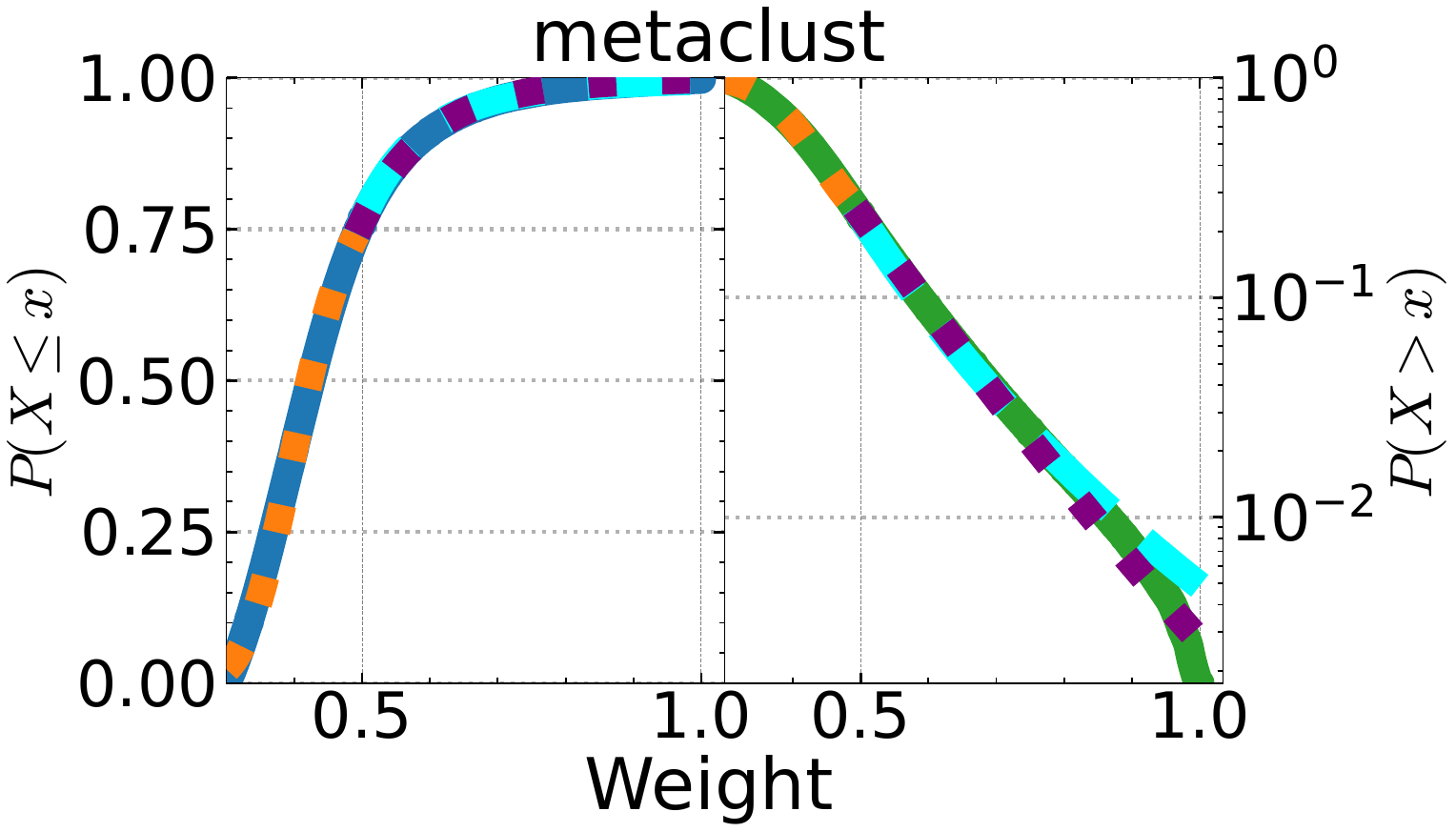}
    \end{subfigure}\hfill
    \caption{Edge weight distributions of the datasets. For each graph, the left panel displays the Cumulative Distribution Function (linear scale) with a superimposed log-normal fit. The right panel shows the Complementary CDF (logarithmic scale) with a power-law and log-normal fit applied to the tail. The x-axis scale varies depending on the specific graph.}
    \label{fig:weighted-datasets-plots}
    \vspace{-3mm}
    \end{figure*}

\subsection{Experimental Setup}
All performance measurements are conducted on a dual-socket AMD~Zen~3 EPYC~7713 processor with 64 cores per socket, for a total of 128 threads, 1TiB DRAM, and 4 NUMA nodes per socket.
SMT is disabled on the processor.
All codebases were compiled with GCC 14.1.0, using the default C++ standard of each respective implementation, while enforcing the \texttt{-O3} and \texttt{-march=native} optimization flags.
Experiments were executed using Linux kernel 4.18.0, and using \texttt{numactl -i all} to interleave memory allocations evenly across all available NUMA nodes, which improves memory bandwidth usage by distributing the graph data.
Threads were pinned to cores during the execution of the experiments. 
The performance measurements are averaged across 22 different start vertices, chosen within the largest connected component, or the largest strongly connected component if the graph is directed.

\section{Characterization of Natural Weights}\label{sec:characterization}

As a starting point for our study, we characterize the empirical edge weight distributions of our graph datasets and contrast them against synthetic weight distributions.
Let $X$ be the random variable representing the possible edge weights. We define 
the cumulative distribution function (CDF) as $F(x)=P(X\leq x)$, and define the complementary CDF (CCDF), as $\overline{F}(x)=1-F(x)=P(X>x)$.

Figure~\ref{fig:weighted-datasets-plots} shows the CDF and CCDF of the weight distribution.
From a first visual inspection we conclude:
\begin{observation}
Synthetic weight distributions are fundamentally different from natural edge weight distributions.
\end{observation}
Indeed, the empirical CDF and CCDF starkly contrast with those expected from synthetic uniform or normal distributions.
Due to space constraints, we only show the weight distributions for some representative graphs; however, all the other graphs follow similar patterns.
In particular, all road graphs span a broad range of weights, with low probability of edges with weight larger than $10^3$; discrete skewed-degree graphs exhibit far fewer unique weight values, due to both their integer nature, and the low probability of values larger than $10$;  biological networks -- generated via sequence similarity algorithms -- exhibit continuous weights strictly bounded between zero and one, resulting in significantly less skewed distributions.

\begin{table*}[t]
\centering
\caption{Statistical characterization of the tail and body for edge weight distributions across evaluated graph datasets. The analysis shows that while the tail behavior varies between power-law (PL) and log-normal (LN) models, the body of the distribution ($x < \xmin$) is universally best described by a log-normal fit.}
\label{tab:fits}
\begin{threeparttable}
\begin{tabular}{lc S[table-format=2.3] S[table-format=5.2] cccc}
\toprule
& \multicolumn{5}{c}{\textbf{Tail}} & \multicolumn{2}{c}{\textbf{Body}} \\
\cmidrule(lr){2-6} \cmidrule(l){7-8}
\textbf{Graph} & \textbf{Best Fit} & $\boldsymbol{\%_{tail}}$ & $\boldsymbol{\xmin}$ & $\boldsymbol{d_{KS}}$ & \textbf{Parameters} & $\boldsymbol{d_{KS}}$ & \textbf{Parameters} \\
\midrule
\multicolumn{8}{c}{\textbf{Road Graphs}} \\
central-america & PL & 0.1\% & 4437.6 & 0.016 & $\alpha = 4.01$ & 0.043 & $\mu = 4.38, \sigma = 1.22$ \\
australia-oceania & LN & 0.3\% & 7378.6 & 0.007 & $\mu = 3.81, \sigma = 1.76$ & 0.043 & $\mu = 3.79, \sigma = 1.66$ \\
south-america & PL & 0.009\% & 21192.4 & 0.009 & $\alpha = 4.08$ & 0.062 & $\mu = 4.45, \sigma = 1.39$ \\
africa & PL & 0.009\% & 28814.6 & 0.011 & $\alpha = 3.68$ & 0.032 & $\mu = 4.55, \sigma = 1.29$ \\
north-america & PL & 0.09\% & 4858.6 & 0.006 & $\alpha = 3.58$ & 0.022 & $\mu = 3.94, \sigma = 1.43$ \\
asia & PL & 0.02\% & 15176.4 & 0.007 & $\alpha = 3.29$ & 0.024 & $\mu = 4.29, \sigma = 1.32$ \\
europe & PL & 0.09\% & 3394.4 & 0.007 & $\alpha = 3.85$ & 0.009 & $\mu = 3.85, \sigma = 1.44$ \\
planet & PL & 0.02\% & 13493.6 & 0.005 & $\alpha = 3.32$ & 0.018 & $\mu = 4.07, \sigma = 1.42$ \\
\midrule
\multicolumn{8}{c}{\textbf{Skewed-degree Graphs}} \\
archaea & LN & 11.6\% & 0.7 & 0.030 & $\mu = -0.29, \sigma = 0.14$ & 0.046 & $\mu = -0.75, \sigma = 0.26$ \\
eukarya & LN & 31.4\% & 0.6 & 0.036 & $\mu = -0.49, \sigma = 0.25$ & 0.048 & $\mu = -0.85, \sigma = 0.20$ \\
isolates-sg1 & LN & 23.1\% & 0.6 & 0.020 & $\mu = -0.78, \sigma = 0.30$ & 0.037 & $\mu = -0.82, \sigma = 0.20$ \\
metaclust & LN & 25.8\% & 0.5 & 0.022 & $\mu = -1.58, \sigma = 0.46$ & 0.034 & $\mu = -0.90, \sigma = 0.16$ \\
mawi\tnote{$\dagger$} & LN & 0.03\% & 241.0 & 0.011 & $\mu = -12.80, \sigma = 5.81$ & 0.349 & $\mu = 0.35, \sigma = 0.51$ \\
coauth-aminer\tnote{$\dagger$} & LN & 0.004\% & 92.0 & 0.026 & $\mu = -44.33, \sigma = 4.18$ & 0.506 & $\mu = 0.13, \sigma = 0.41$ \\
coauth-mag\tnote{$\dagger$} & LN & 0.03\% & 62.0 & 0.036 & $\mu = -123.18, \sigma = 7.30$ & 0.455 & $\mu = 0.25, \sigma = 0.54$ \\
colisten-spotify\tnote{$\dagger$} & LN & 0.7\% & 92.0 & 0.010 & $\mu = -27.46, \sigma = 5.86$ & 0.349 & $\mu = 0.53, \sigma = 0.84$ \\
moliere & PL & 15.4\% & 0.1 & 0.028 & $\alpha = 3.10$ & 0.035 & $\mu = -4.05, \sigma = 0.87$ \\
\bottomrule
\end{tabular}%
\begin{tablenotes}[para, flushleft]\footnotesize
  \item \textbf{PL}: Power-law;
  \item \textbf{LN}: Log-normal;
  \item $\boldsymbol{\%_{tail}}$: Percentage of edge weights $x \geq \xmin$;
  \item $\boldsymbol{\xmin}$: Tail starting point;
  \item $\boldsymbol{d_{KS}}$: Kolmogorov-Smirnov distance;
  \item[$\dagger$] The graph has integer weights.
\end{tablenotes}
\end{threeparttable}%
\end{table*}

\subsection{Fitting Weight Distributions}

To better understand the properties of real-world graphs, we systematically characterize their empirical edge weight distributions.
By doing so, we aim to inform future benchmark design and provide critical context for the performance analysis in Section~\ref{sec:performance}.
The CDF and the log-log CCDF provide immediate visual intuition regarding the structure of the data; in particular, they highlight the clear distinction between the body and the (heavy) tail of the distributions, which must be modeled independently.

To characterize both the body and the tail of the weight distributions, we adopt the statistical framework established by Clauset et al.~\cite{clausetPowerlawDistributionsEmpirical2009}, which
has been previously utilized to analyze the degree distributions of various real-world graphs~\cite{garciaSocialResilienceOnline2013, meuselGraphStructureWeb2014, meuselGraphStructureWeb2015, broidoScalefreeNetworksAre2019}.
Furthermore, although the method was originally formulated to analyze tail behavior, its principles can be adapted to rigorously characterize the body of the distribution as well.
The generalized pipeline consists of three steps: 

\begin{enumerate}
    \item \textit{\textbf{Parameter Estimation}}: We estimate the parameters of the hypothesized distribution using a Maximum Likelihood Estimator (MLE)~\cite{casellaStatisticalInference2024, clausetPowerlawDistributionsEmpirical2009}, when it provides a closed-form solution, or otherwise numerical optimization.
    The maximum likelihood method estimates the parameters of an hypothesized distribution given the empirical data.
    This is achieved by maximizing a likelihood function, systematically calculating the distribution parameters that render the empirical data mathematically most probable.
    
    All the natural weight distributions that we studied are best modeled using different distributions for their body and their tail.
    The optimal boundary separating the body from the tail, $\xmin$, is established by finding the value that minimizes the Kolmogorov-Smirnov (KS) distance~\cite{pressNumericalRecipes2nd1992} between the empirical data and the best-fit model for the tail. 
    Edge weight samples larger than $\xmin$ are subsequently used to fit the left-truncated distribution for the tail, and the samples smaller than $\xmin$ are used to fit the body of the distribution.

    \item \textit{\textbf{Goodness-of-Fit}}: MLE identifies the best-fitting parameter values, but the procedure does not prove that the shape of the distribution function is appropriate. To test the plausibility of the distribution shape, we use
    a semi-parametric bootstrapping procedure~\cite{efronIntroductionBootstrap1994}. 
    We generate random datasets following the proposed edge weight distribution.
    Each dataset consists of a large number of randomly generated edge weights which we independently fit to the proposed distribution function
    and calculate its KS distance. 
    The $p$-value is formally defined as the fraction of synthetic datasets whose KS distance is greater than or equal to the empirical KS distance. 
    A high $p$-value indicates that the empirical deviation from the theoretical model could reasonably be attributed to random statistical fluctuations, whereas a low $p$-value (typically $p \leq 0.1$) signifies that the deviation is too large to be random, resulting in the rejection of the hypothesis.
    
    \item \textit{\textbf{Alternative Hypotheses}}: Even if a specific distribution passes the goodness-of-fit test, another model might offer a superior fit. 
    Therefore, we perform a likelihood-ratio test to quantitatively compare the primary hypothesized model against alternative candidate distributions appropriate for that data domain.
\end{enumerate}

Applying this methodology to the scale of our datasets presented a significant computational challenge.
Standard software tools, such as the Python \texttt{powerlaw} package~\cite{alstottPowerlawPythonPackage2014} or the \texttt{plfit} implementation, proved computationally prohibitive at this scale, or otherwise required random sampling that could compromise the integrity of the tail analysis. 
To overcome this, we developed a parallel C++ implementation of the Clauset et al. methodology.
Rather than processing raw weight arrays, our implementation  operates on an exact histogram of the empirical weight data, precomputes intermediate results, and tests the KS distance of the $x_{min}$ candidates in parallel.

\begin{figure*}[t]
    \centering
    \begin{subfigure}[b]{\textwidth}
        \centering
        \includegraphics[width=0.7\textwidth]{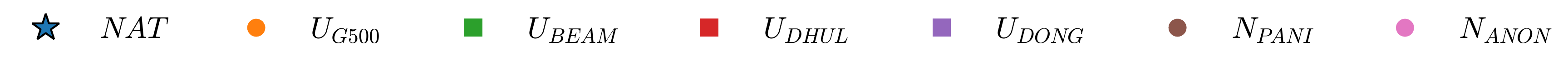}
    \end{subfigure}
    \begin{subfigure}[b]{0.23\textwidth}
        \centering
        \includegraphics[width=\textwidth]{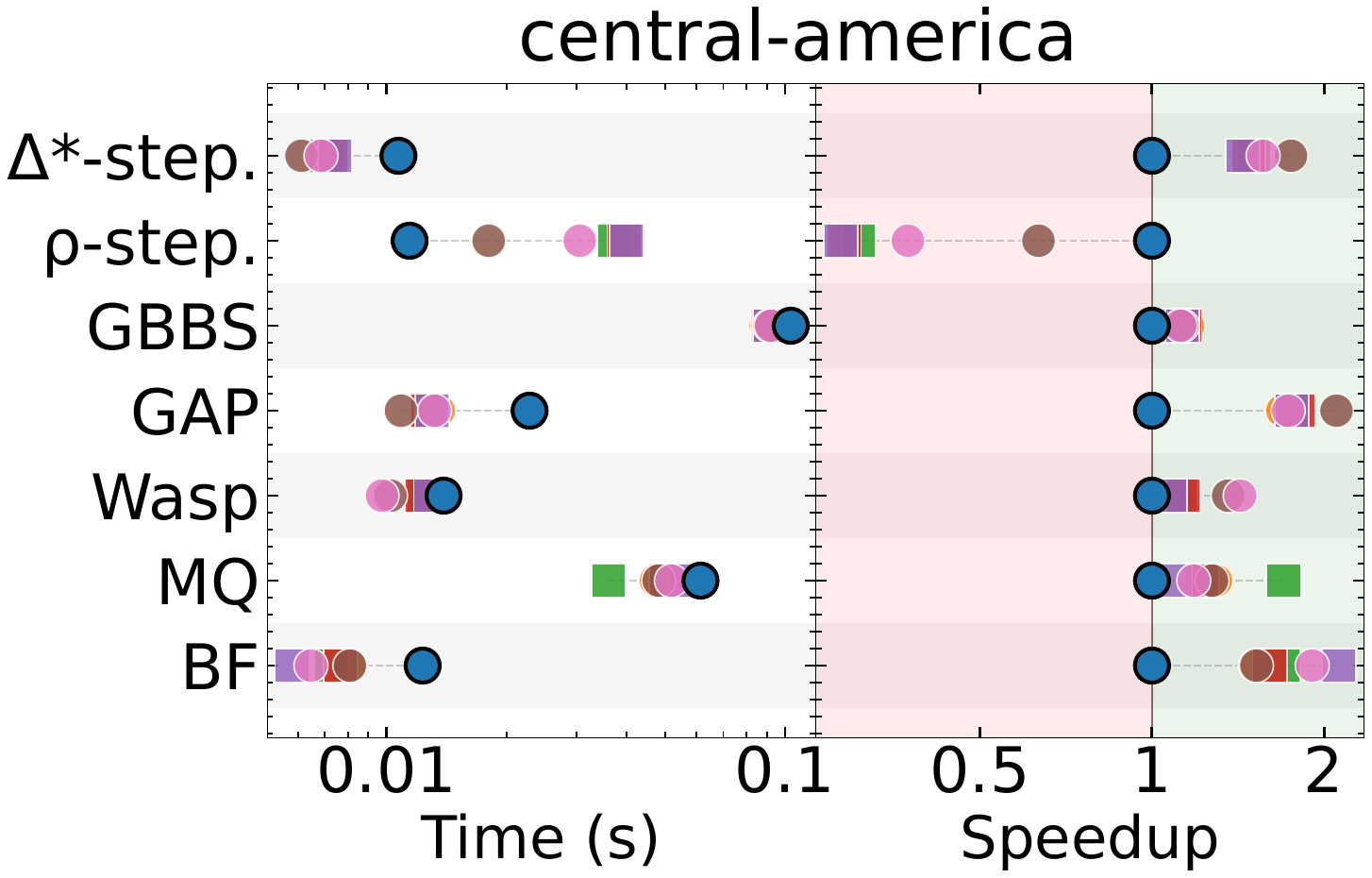}
    \end{subfigure}\hfill
    \begin{subfigure}[b]{0.23\textwidth}
        \centering
        \includegraphics[width=\textwidth]{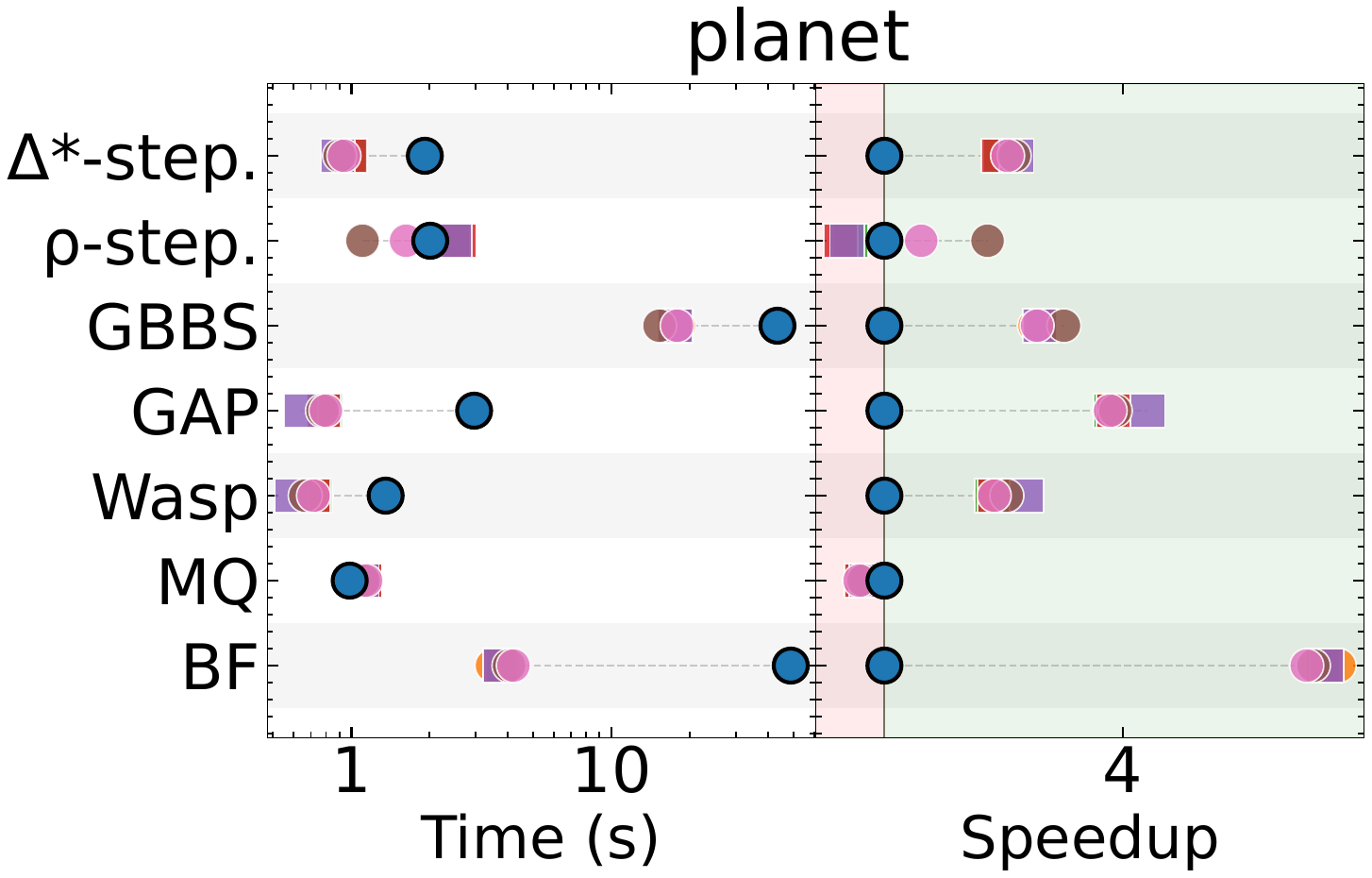}
    \end{subfigure}\hfill
    \begin{subfigure}[b]{0.23\textwidth}
        \centering
        \includegraphics[width=\textwidth]{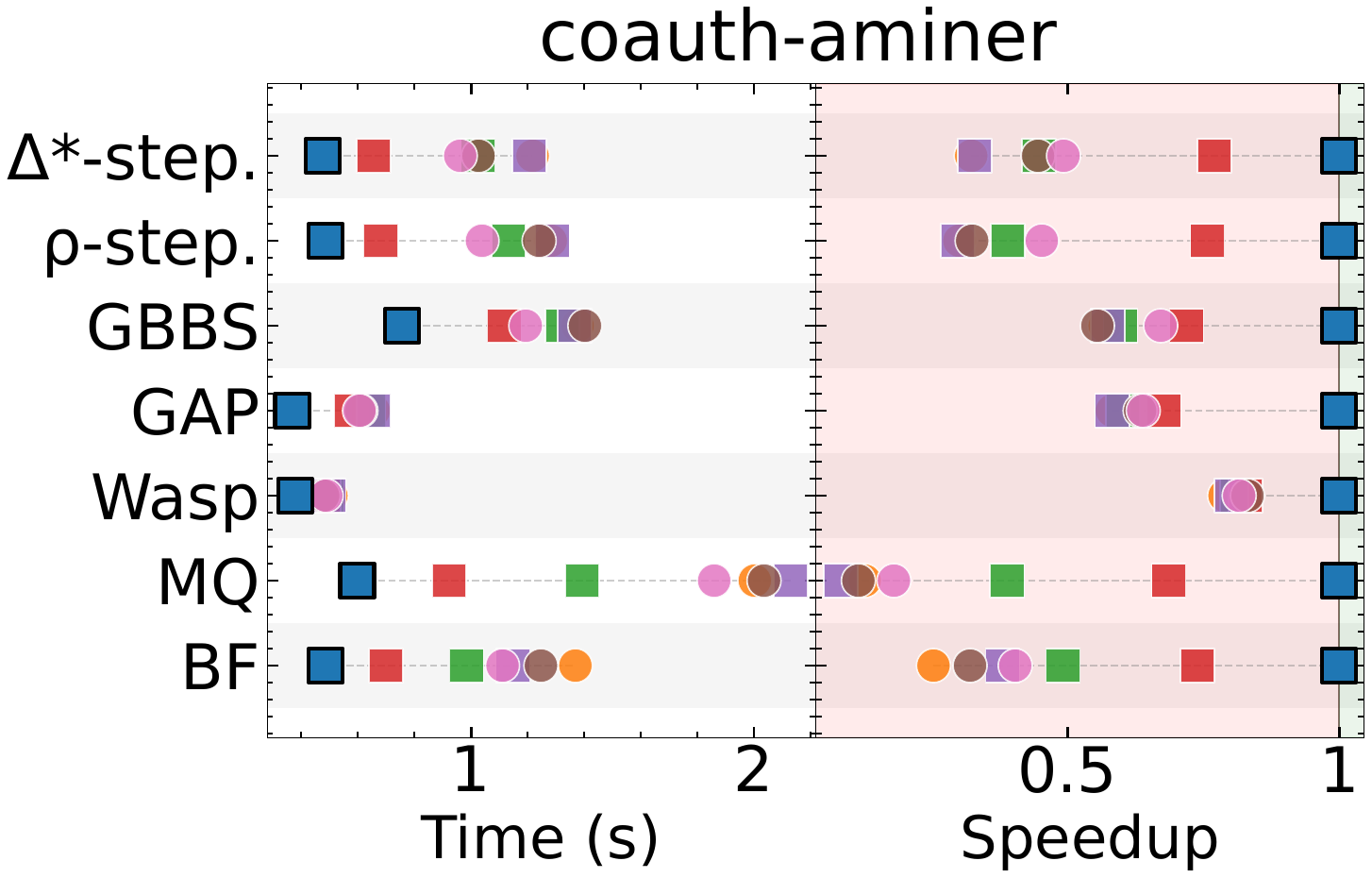}
    \end{subfigure}\hfill
    \begin{subfigure}[b]{0.23\textwidth}
        \centering
        \includegraphics[width=\textwidth]{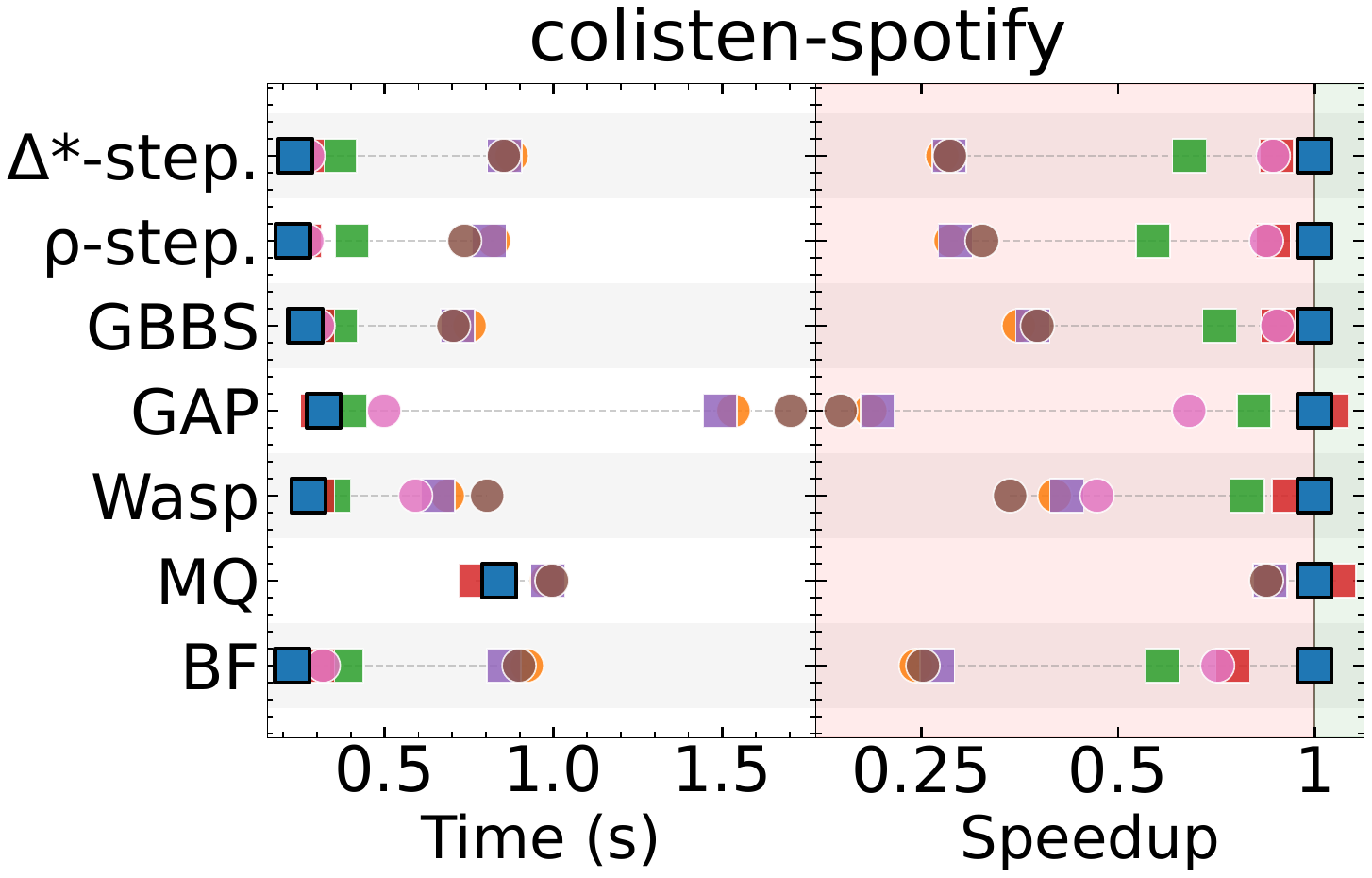}
    \end{subfigure}\hfill
    \begin{subfigure}[b]{0.23\textwidth}
        \centering
        \includegraphics[width=\textwidth]{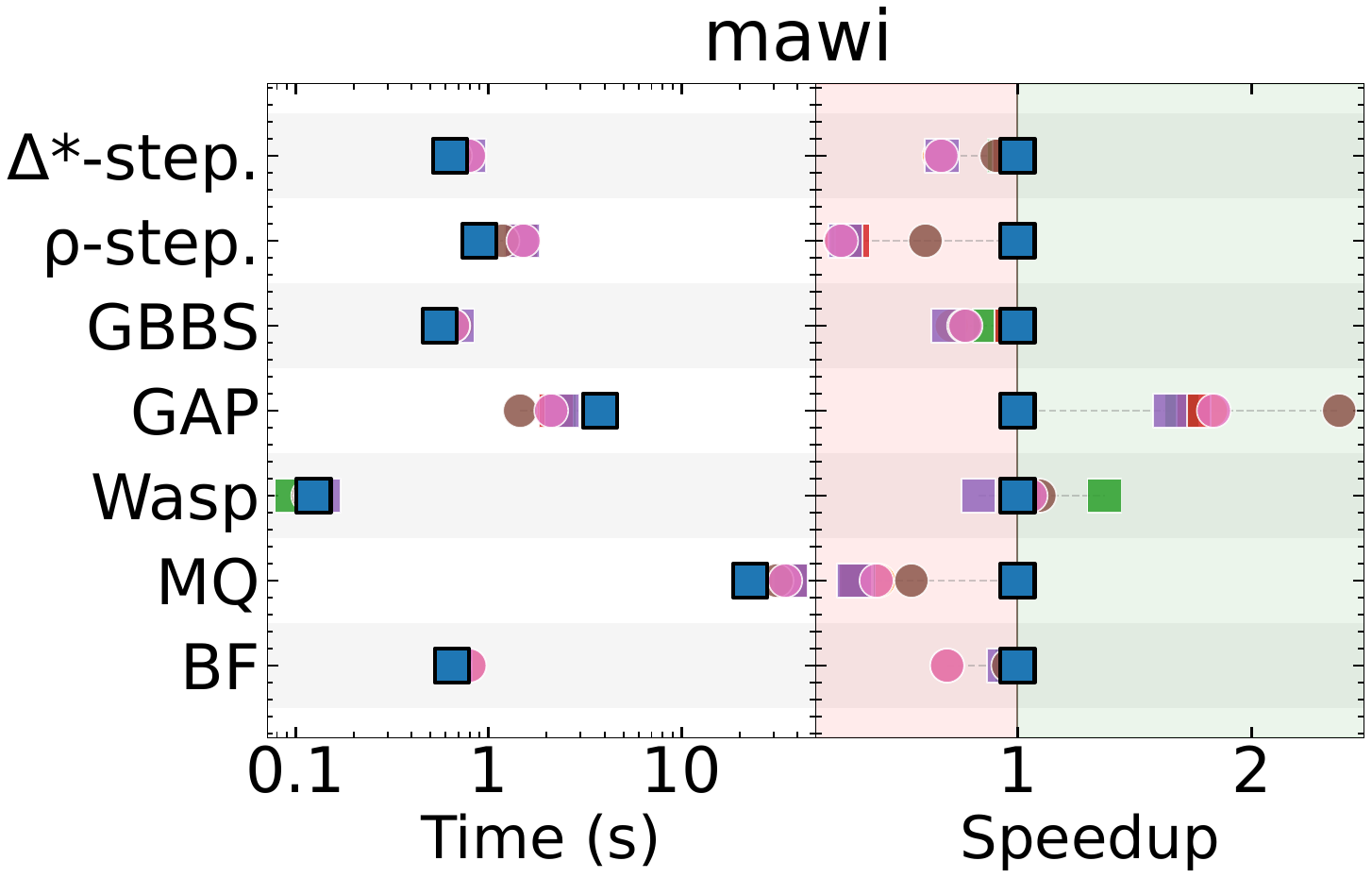}
    \end{subfigure}\hfill
    \begin{subfigure}[b]{0.23\textwidth}
        \centering
        \includegraphics[width=\textwidth]{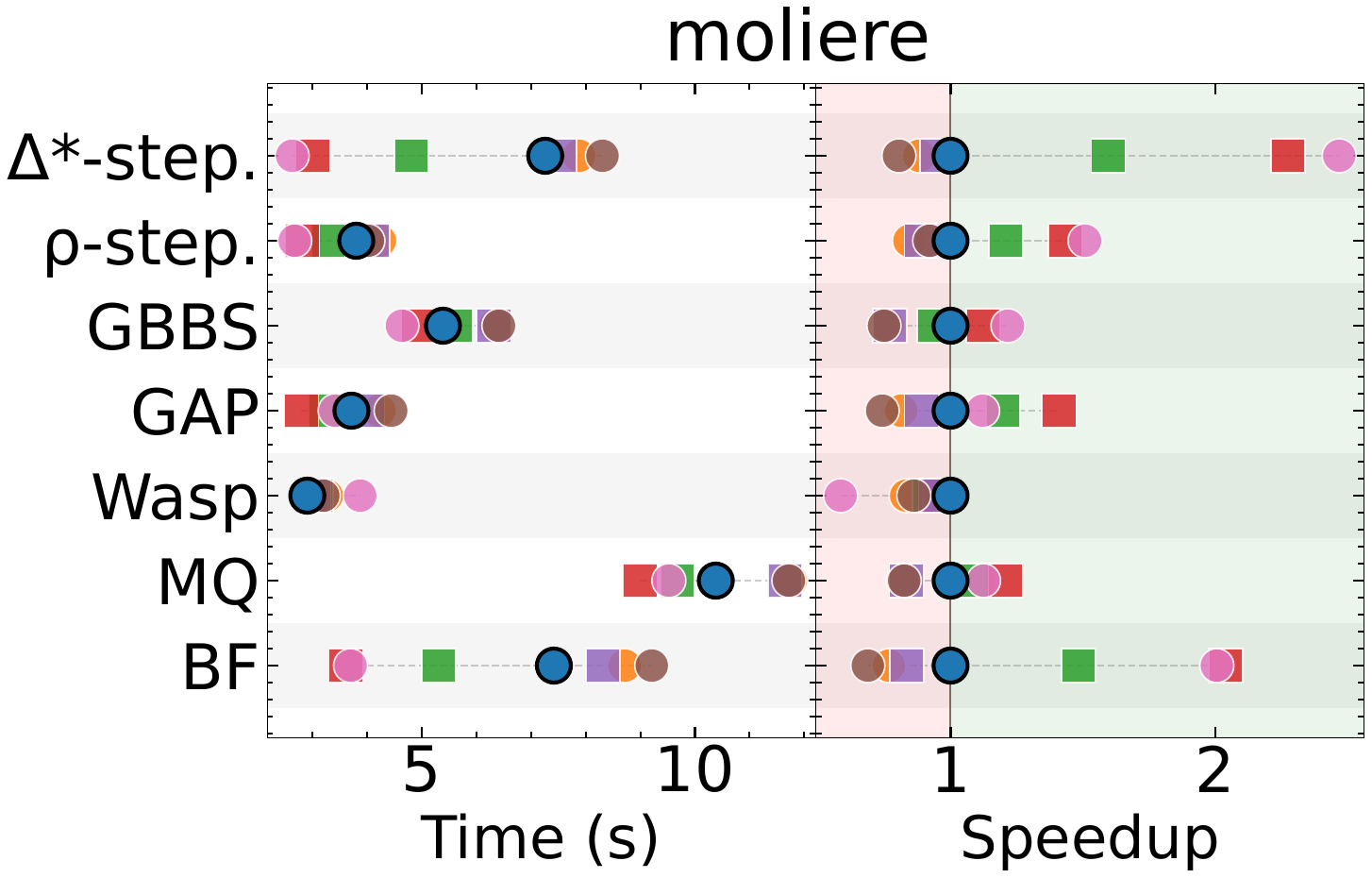}
    \end{subfigure}\hfill
    \begin{subfigure}[b]{0.23\textwidth}
        \centering
        \includegraphics[width=\textwidth]{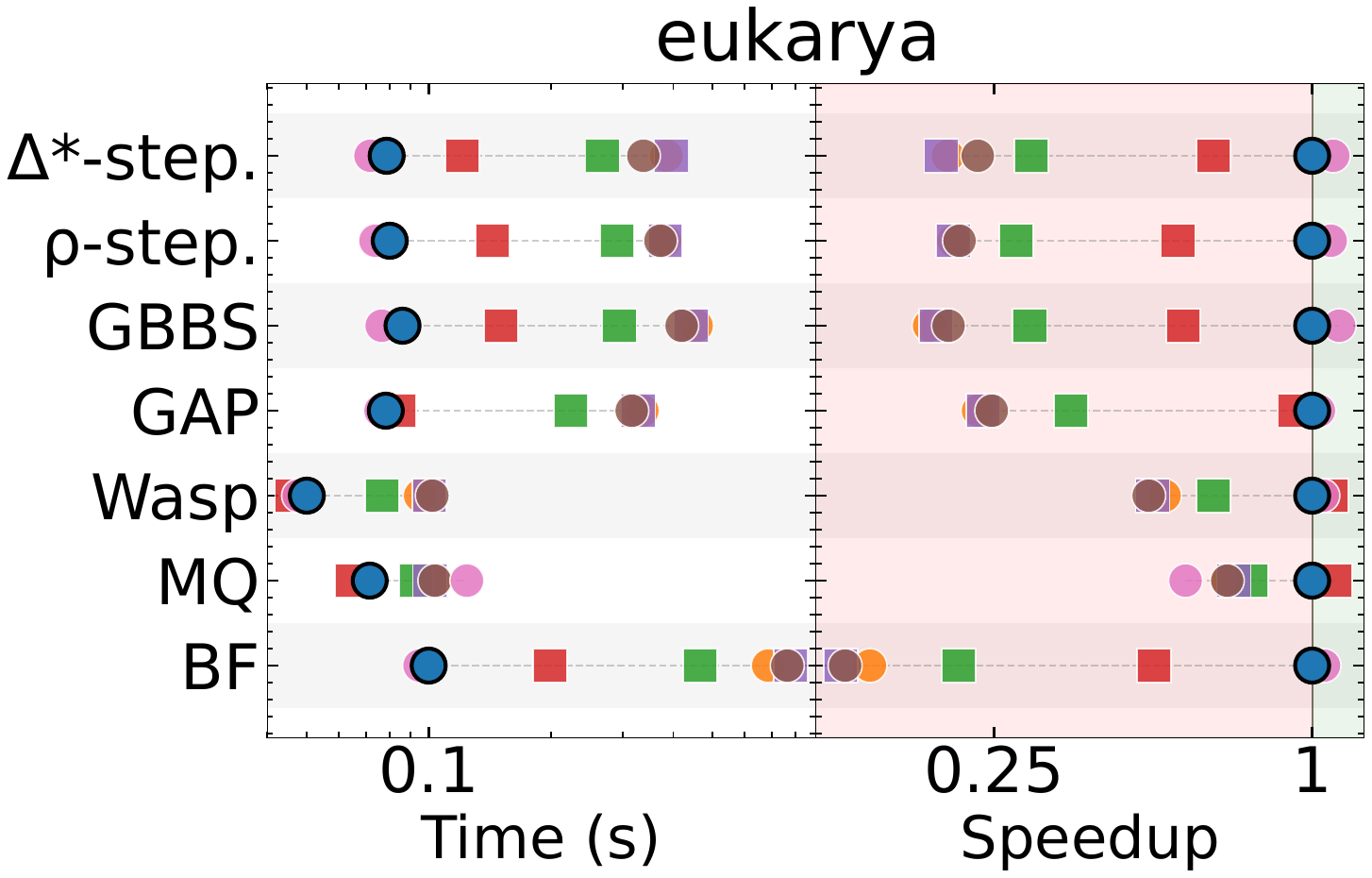}
    \end{subfigure}\hfill
    \begin{subfigure}[b]{0.23\textwidth}
        \centering
        \includegraphics[width=\textwidth]{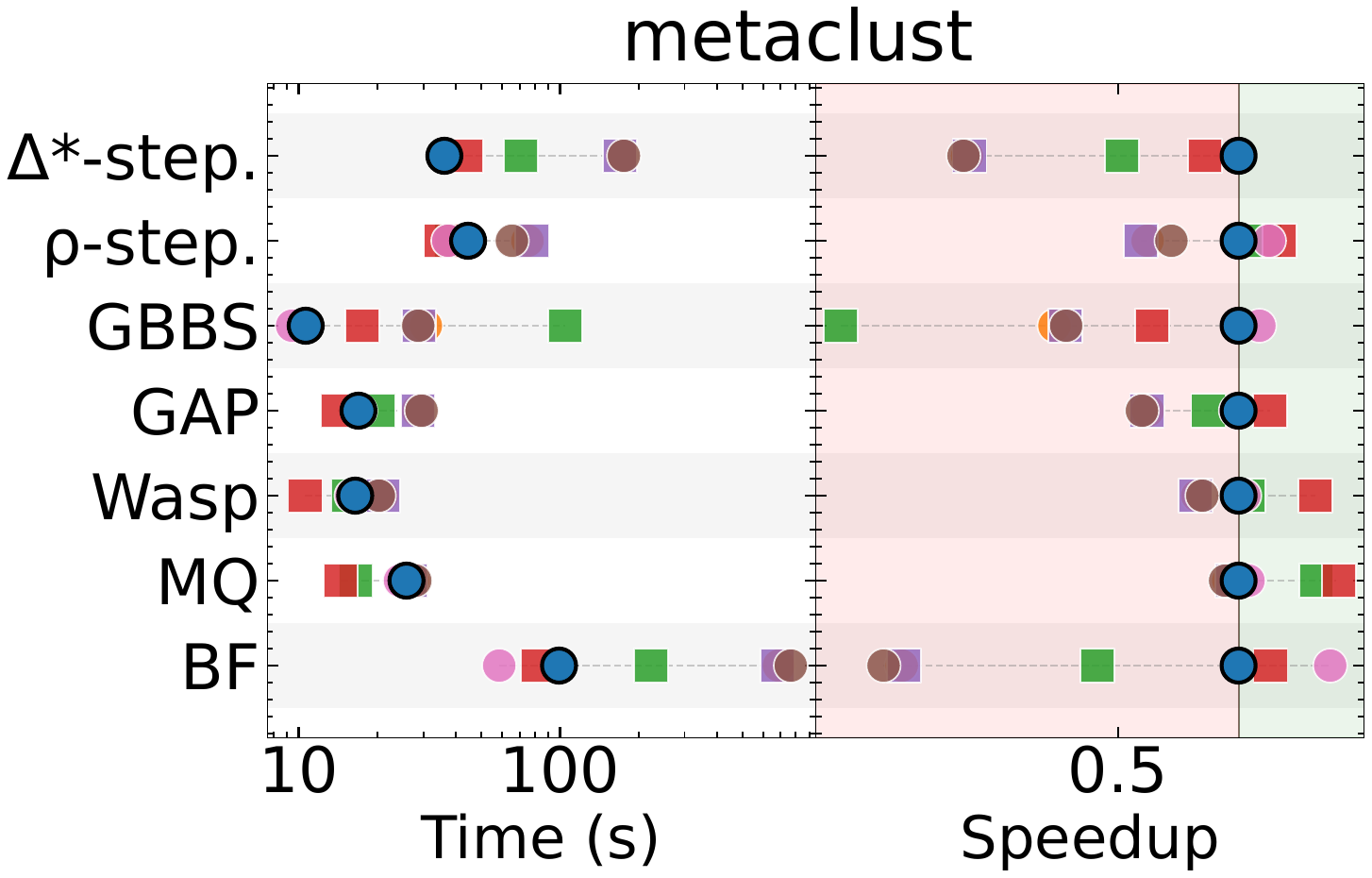}
    \end{subfigure}\hfill
    \caption{Performance of SSSP implementations on different graphs for each weight distribution. The left panel indicates the execution time, the right panel shows the speedup of the synthetic weight distributions against the natural weight distribution. Integer weights are indicated with a square symbol, while float weights are indicated with a circle symbol.}
    \label{fig:performance}
\end{figure*}

\subsection{Analysis of the Distribution's Tail}
In the tail analysis, we evaluate the power-law hypothesis against two alternative heavy-tailed models: a power law with an exponential cutoff, and a log-normal distribution.
Because our analysis focuses strictly on the tail domain ($x \geq x_{min}$), we fit a left-truncated log-normal distribution to the data.
To perform the goodness-of-fit test, we evaluated the power-law hypothesis using 2500 synthetic datasets -- as suggested by Clauset et al.~\cite{clausetPowerlawDistributionsEmpirical2009} to achieve a $p$-value precision of $\pm 0.01$. 
For the vast majority of our datasets, this test resulted in a $p$-value of $p < 0.01$, therefore rejecting the power-law hypothesis. 
We observed statistically plausible fits ($p > 0.1$) for only two networks: \texttt{africa} ($p=0.7336$) and \texttt{south-america} ($p=0.9792$).
As shown in statistical literature~\cite{linResearchCommentaryToo2013, demidenkoPValueYouCant2016, broidoScalefreeNetworksAre2019}, when evaluating hypotheses for massive datasets, $p$-values are known to rapidly approach zero, as minor empirical fluctuations become statistically significant enough to reject the strict theoretical model.
Indeed, the power-law hypothesis was validated only in such datasets where the tail encompasses a very small fraction (roughly $0.009\%$) of the total edges.
To determine which distribution models the tail most accurately, Table~\ref{tab:fits} reports the superior model determined by the likelihood-ratio test. 
We observe that the log-normal hypothesis is a better tail fit for more than half of the evaluated datasets.
This conclusion is further supported visually in Figure~\ref{fig:weighted-datasets-plots}, where the power-law and log-normal tail fits are indicated by cyan and purple dashed lines, respectively.

\begin{observation}
    The tail of the edge weight distribution does not always follow a power-law distribution.
\end{observation}

\subsection{Analysis of the Distribution's Body}
The body of the weight distribution is where the majority of data points reside; for this reason, characterizing the behavior of the body can provide insights useful to replicate natural distributions.
Visual inspection of the CDF in Figure~\ref{fig:weighted-datasets-plots} suggests a log-normal fit for the data.
To confirm this we fit a right-truncated log-normal distribution, setting the upper bound to the $\xmin$ value previously identified for the tail.
Running a full bootstrap goodness-of-fit test on the distribution body was computationally prohibitive.
Furthermore, because the empirical body accounts for $68\%$ to $99.991\%$ of the total data depending on the dataset, any theoretical fit would be rejected by the test due to the same large-sample artifacts discussed in the tail analysis~\cite{linResearchCommentaryToo2013, demidenkoPValueYouCant2016}.
Instead, we rely on the likelihood-ratio test to compare the log-normal hypothesis against two alternative models: a right-truncated exponential distribution and a right-truncated Weibull distribution.
In all cases, the likelihood-ratio test confirmed that the log-normal distribution provides a statistically better fit.
Figure~\ref{fig:weighted-datasets-plots} shows the log-normal body fit with an orange dotted line, while Table~\ref{tab:fits} indicates the KS distance of the fit and the parameters of the right-truncated log-normal distribution.
We observe that graphs belonging to the same structural class exhibit similar distribution parameters, which is reflected in the similar shapes in Figure~\ref{fig:weighted-datasets-plots}.
Integer skewed-degree graphs proved particularly difficult to fit, as evidenced by shown by their larger KS distances and the shape of the curve in the CCDF, which exhibits an initial overestimation of the probability before flattening out as it approaches $\xmin$.

\begin{observation}
    The log-normal distribution provides a robust fit for the body of empirical edge weight distributions.
\end{observation}

\section{Performance Analysis}\label{sec:performance}
To evaluate the impact of weight distributions on execution time, we determine the optimal $\Delta$ for each combination of SSSP implementation, graph, and weight distribution.
We then compare these best-performing configurations.

\subsection{Performance Impact of Weight Distribution}
Figure~\ref{fig:performance} illustrates the execution time of the SSSP implementations across eight representative graphs, highlighting the relative speedup of synthetic weight distributions compared to the natural weight baseline.

\paragraph{\textbf{Road Graphs}}
For road graphs, we observe that most SSSP implementations execute significantly faster when using synthetic weights rather than natural weights.
A notable exception is \rhostep, which exhibits high variability; it experiences up to a roughly $2\times$ slowdown on \texttt{central-america} and shows similar, though less pronounced, slowdowns on other road graphs. 
The performance gap is particularly significant for the parallel Bellman-Ford algorithm:
with naturally weighted road graphs, Bellman-Ford suffers from very high and variable execution times depending on the traversal's source vertex, whereas synthetic weights artificially mask this inefficiency, resulting in massive apparent speedups.
In fact, for certain graphs and synthetic weight distributions -- such as \texttt{africa}, \texttt{australia-oceania}, and \texttt{central-america} -- Bellman-Ford is among the most efficient algorithms evaluated.

\begin{figure*}[t]
    \centering
    \begin{subfigure}[b]{0.52\textwidth}
        \centering
        \includegraphics[width=\textwidth]{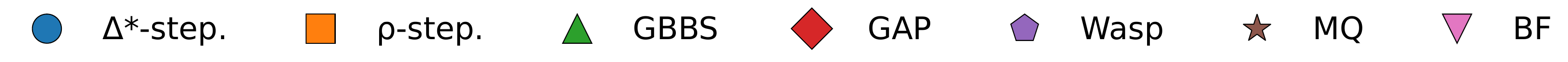}
        \includegraphics[width=\textwidth]{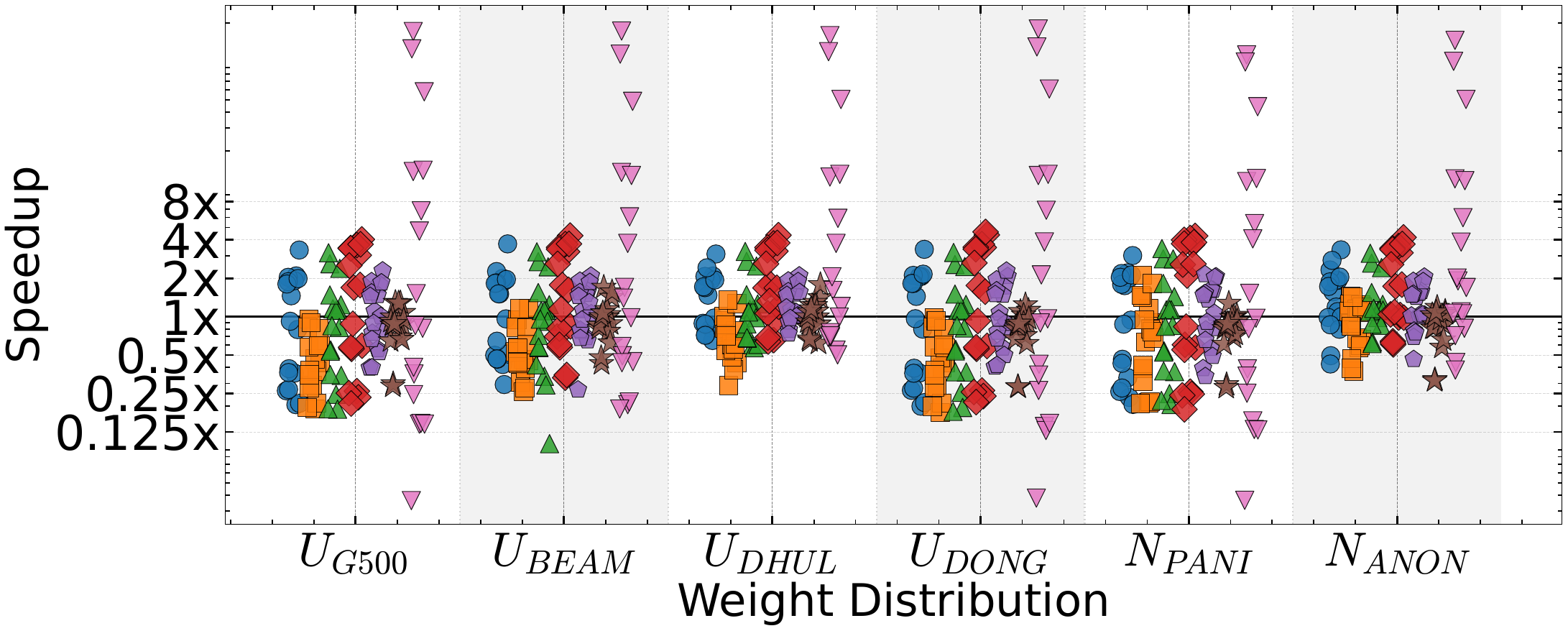}
        \caption{Relative performance of the implementation across different distributions compared to the natural weights baseline.}
        \label{fig:speedup-scatter}
    \end{subfigure}\hfill
    \begin{subfigure}[b]{0.47\textwidth}
        \centering
        \includegraphics[width=\textwidth]{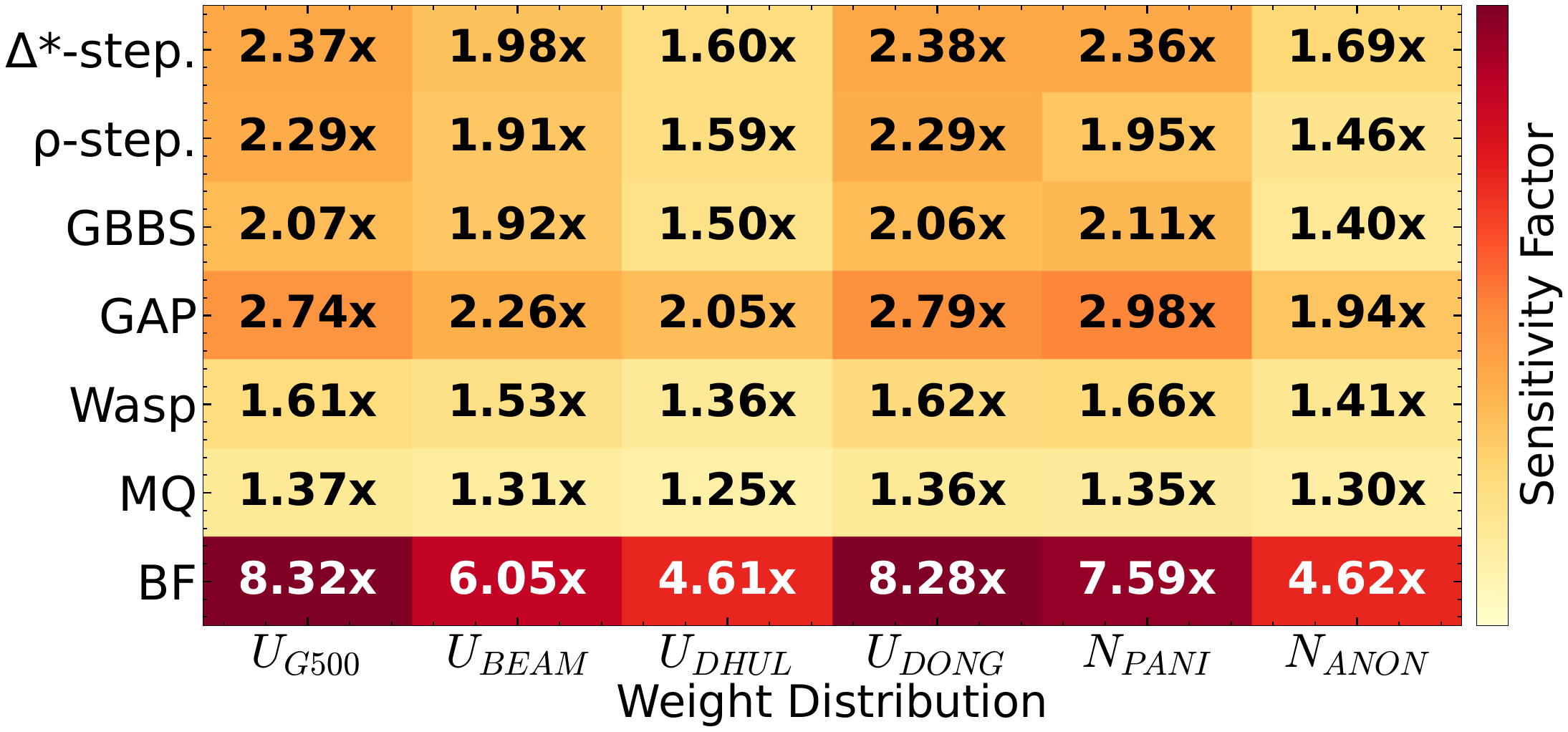}
        \caption{Sensitivity of each implementation to different weight distribution showing the relative performance variability.}
        \label{fig:speedup-heatmap}
    \end{subfigure}
    \caption{Sensitivity analysis of the parallel SSSP implementations to synthetic weight distributions.}
    \label{fig:speedup}
\end{figure*}

\paragraph{\textbf{Skewed-degree Graphs}}
In contrast to road graphs, synthetic weight distributions generally degrade performance on skewed-degree graphs.
Substituting natural weights with synthetic counterparts consistently shifts the execution into the slowdown region.
In several cases, synthetic distributions induce up to a $4\times$ slowdown compared to the natural weights (visible prominently in implementations like GAP, Wasp, and MQ on \texttt{coauth-aminer}).

\begin{observation}
    Synthetic weight distributions misrepresent real-world SSSP performance, artificially inflating execution speeds on large-diameter graphs while exhibiting slowdowns and high-variability on skewed-degree topologies.
\end{observation}

\subsection{Performance Sensitivity}
Figure~\ref{fig:speedup-scatter} shows the relative speedup obtained when substituting natural weights with synthetic distributions, where each data point corresponds to a distinct graph dataset.
We summarize the results in Figure~\ref{fig:speedup-heatmap} using a sensitivity factor.
This metric is computed as the geometric mean of the absolute logarithmic speedups measured between synthetic distributions and the natural weight distribution.
By taking the absolute value, the metric ensures that both performance improvements and degradations are symmetrically captured, preventing opposing fluctuations from canceling each other out.
Overall, synthetic weights consistently fail to mirror real-world execution, inducing both severe artificial speedups and massive slowdowns.
The Bellman-Ford algorithm is particularly sensitive to the weight distribution with speedups as high as $180\times$ or as low as $0.04\times$.
\dstarstep, \rhostep, GBBS, and GAP exhibit similar overall sensitivity factors in the heatmap; however, the scatter plot reveals distinct underlying behaviors.
Specifically, \rhostep predominantly suffers slowdowns across all synthetic weight distributions, whereas the other three implementations display a more balanced variance around the natural weight baseline.

\begin{figure}[t]
    \centering
    \includegraphics[width=.85\linewidth]{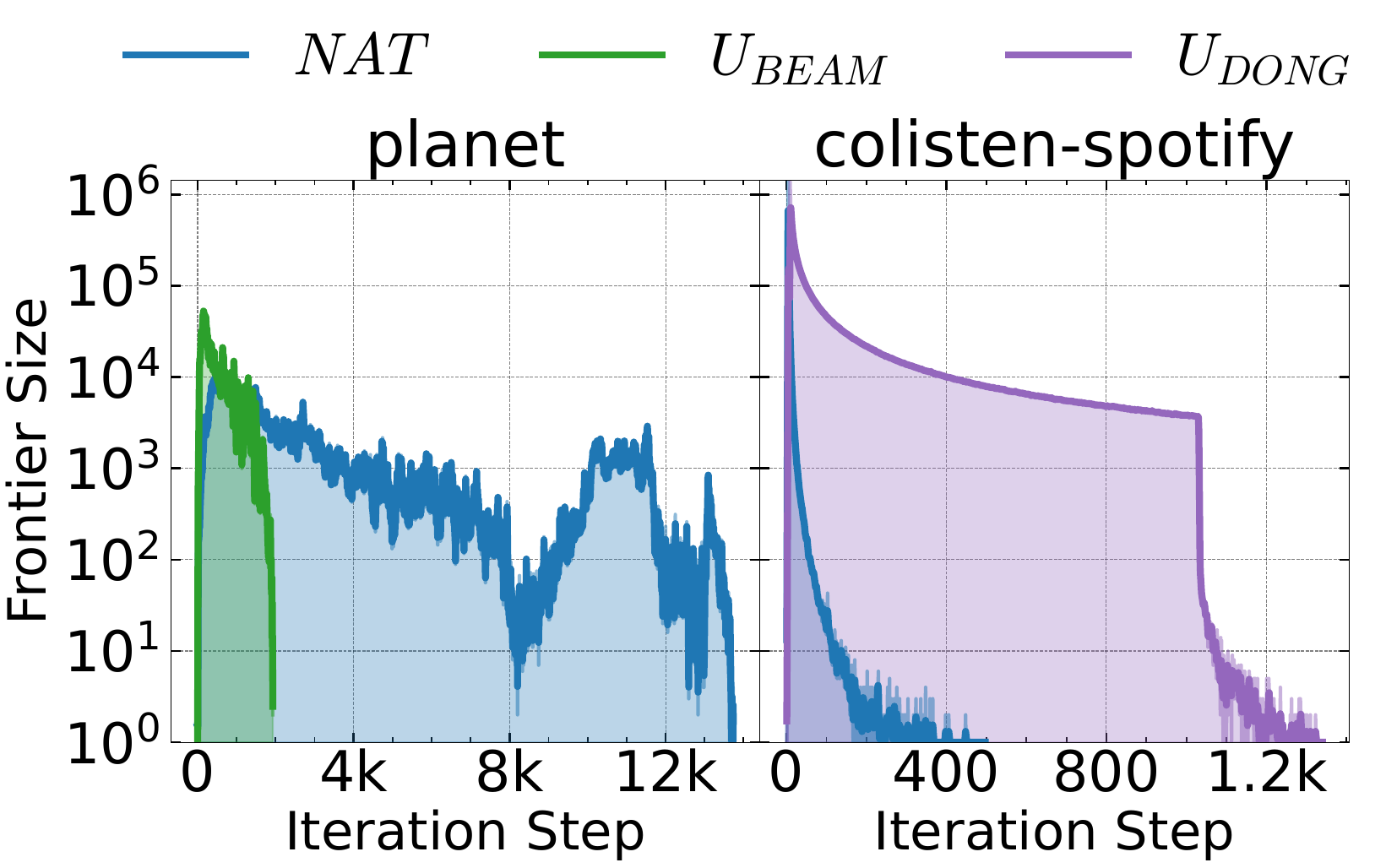}
    \caption{Size of the frontier (in vertices) in each synchronous execution step in GAP SSSP.}
    \label{fig:frontier-size}
\end{figure}

We investigate the mechanics underlying these performance variations by plotting
the frontier size of GAP (a synchronous implementation) on the \texttt{planet} and \texttt{colisten-spotify} datasets in Figure~\ref{fig:frontier-size}.
On the \texttt{planet} graph, executing with natural weights requires $7\times$ more iteration steps compared to the $U_{BEAM}$ distribution, while the total number of edge relaxations is only 11\% larger.
Different edge weights can thus cause a reshuffling of active vertices to different synchronous iterations.
This occurs, e.g., when a large edge weight placed in a critical position assigns a large tentative distance to a high-degree vertex, postponing the processing of a large number of neighboring vertices for several iterations.
As such, edge weights dictate the required number of BSP steps, independently of topology.
On the \texttt{colisten-spotify} graph, the $U_{DONG}$ distribution incurs a $4\times$ slowdown relative to natural weights.
Here, the synthetic weights trigger both a $3.4\times$ increase in relaxations and a $2.7\times$ increase in BSP steps, proving that weight distributions can simultaneously inflate both the computational work and the number of synchronous iterations.
Such a scenario could be caused
by the presence of parallel paths, where the order of processing paths enforces repeated re-activation of the same vertices. 

In contrast, Wasp and the MultiQueue (MQ) exhibit the lowest sensitivity to weight distributions.
This robustness is visually confirmed by their tightly clustered data points in the scatter plot -- disrupted only by few outliers -- and their consistently low sensitivity scores across all evaluated distributions (Figure~\ref{fig:speedup-scatter}).
We attribute the lower sensitivity in these two implementations to asynchrony.
Asynchronous algorithms progress the computation by processing the next best vertex, irrespective of their tentative distance compared to an artificial $\Delta$-based distance.

\begin{observation}
Edge weights dictate both computational work and the distribution of work over synchronous steps, causing performance instability for synchronous algorithms. 
Asynchronous algorithms absorb variations in edge weights, yielding more robust performance.
\end{observation}

\section{\texorpdfstring{$\Delta$}{Delta}-tuning Analysis}\label{sec:tuning}
For SSSP algorithms based on the $\Delta$-stepping paradigm, the underlying edge weight distribution inherently influences the optimal tuning of the $\Delta$ parameter.
To systematically evaluate this impact, we profile the execution time of each implementation across a wide sweep of $\Delta$ values.
Specifically, we test power-of-two increments ranging from $2^{-20}$ to $2^{25}$.
For natural and synthetic integer distributions, we restrict this sweep to integer values ($\Delta \geq 1$).

\subsection{Tuning Penalty}

\begin{figure}[t]
    \centering
    \includegraphics[width=.95\linewidth]{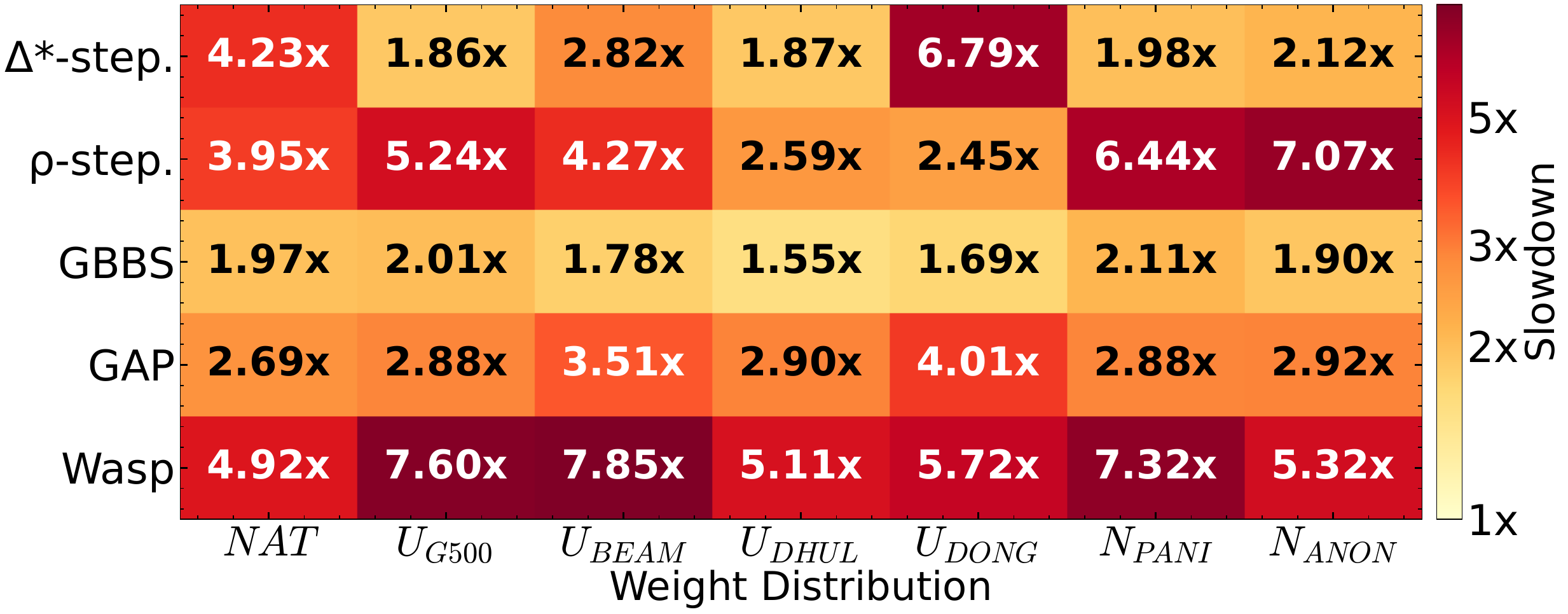}
    \caption{Aggregated $\Delta$ mis-tuning penalty across graphs. Values approaching $1.0\times$ indicate robustness to $\Delta$; darker cells reveal severe sensitivity to parameter tuning.}
    \label{fig:tuning-penalty}
\end{figure}

Figure~\ref{fig:tuning-penalty} illustrates the aggregated $\Delta$ mis-tuning penalty -- calculated as the geometric mean of the slowdowns relative to the optimal $\Delta$ configuration -- across all graphs for each weight distribution.
We observe that Wasp suffers from the highest average mis-tuning penalty,  demonstrating severe sensitivity to suboptimal parameters.
This is followed by \rhostep, which exhibits high penalties particularly on the $U_{G500}$, $N_{PANI}$, and $N_{ANON}$ distributions.
Conversely, GBBS proves to be the most robust framework regarding parameter tuning, exhibiting the lowest average slowdowns across all evaluated distributions.

Several methodological caveats are necessary to contextualize these results.
First, GBBS failed to produce correct results for $\Delta < 1$ on naturally weighted road networks; these invalid runs are excluded.
Second, because heavily mis-tuned $\Delta$ values frequently caused execution times to exceed our strict ten-minute job limit (a threshold sufficient to complete all trials with an optimally tuned $\Delta$), not every implementation successfully completed the entire parameter sweep.
To ensure a fair comparison, the aggregated penalties shown in Figure~\ref{fig:tuning-penalty} are computed exclusively over the intersection of $\Delta$ values that successfully executed across all implementations for a given graph.
This intersection yields an average of 22 valid $\Delta$ data points per configuration.

\begin{observation}
    Mis-tuning slowdown is present across all distributions, highlighting how $\Delta$-tuning is crucial for achieving peak  performance. However, the severity of this penalty is fundamentally dictated by algorithmic design, with some frameworks proving highly resilient while others suffer drastic performance degradation when suboptimally configured.
\end{observation}

\subsection{Tuning Behavior}
\begin{figure*}[t]
    \centering
    \begin{subfigure}[b]{\textwidth}
        \centering
        \includegraphics[width=0.5\textwidth]{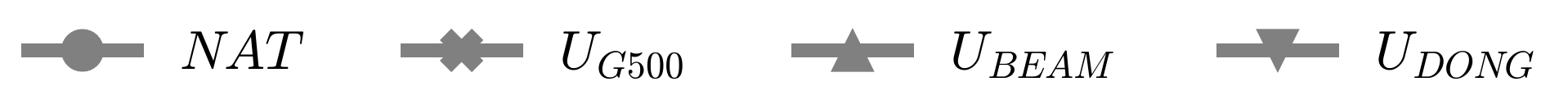}\hfill
        
        \includegraphics[width=0.7\textwidth]{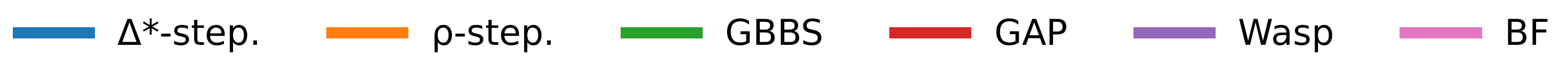}
    \end{subfigure}
    \begin{subfigure}[b]{0.49\textwidth}
        \centering
        \includegraphics[width=.87\textwidth]{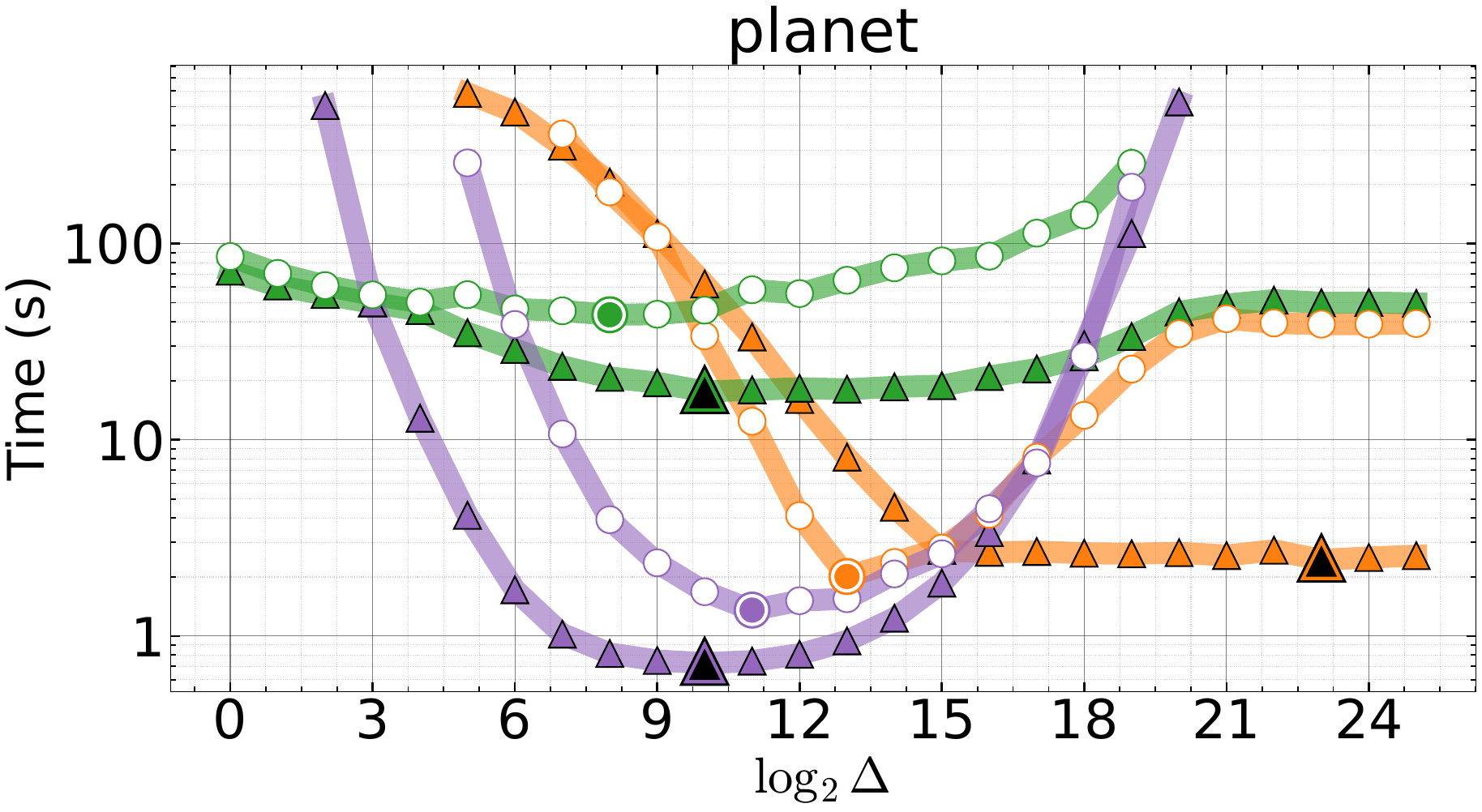}
    \end{subfigure}\hfill
    \begin{subfigure}[b]{0.49\textwidth}
        \centering
        \includegraphics[width=.87\textwidth]{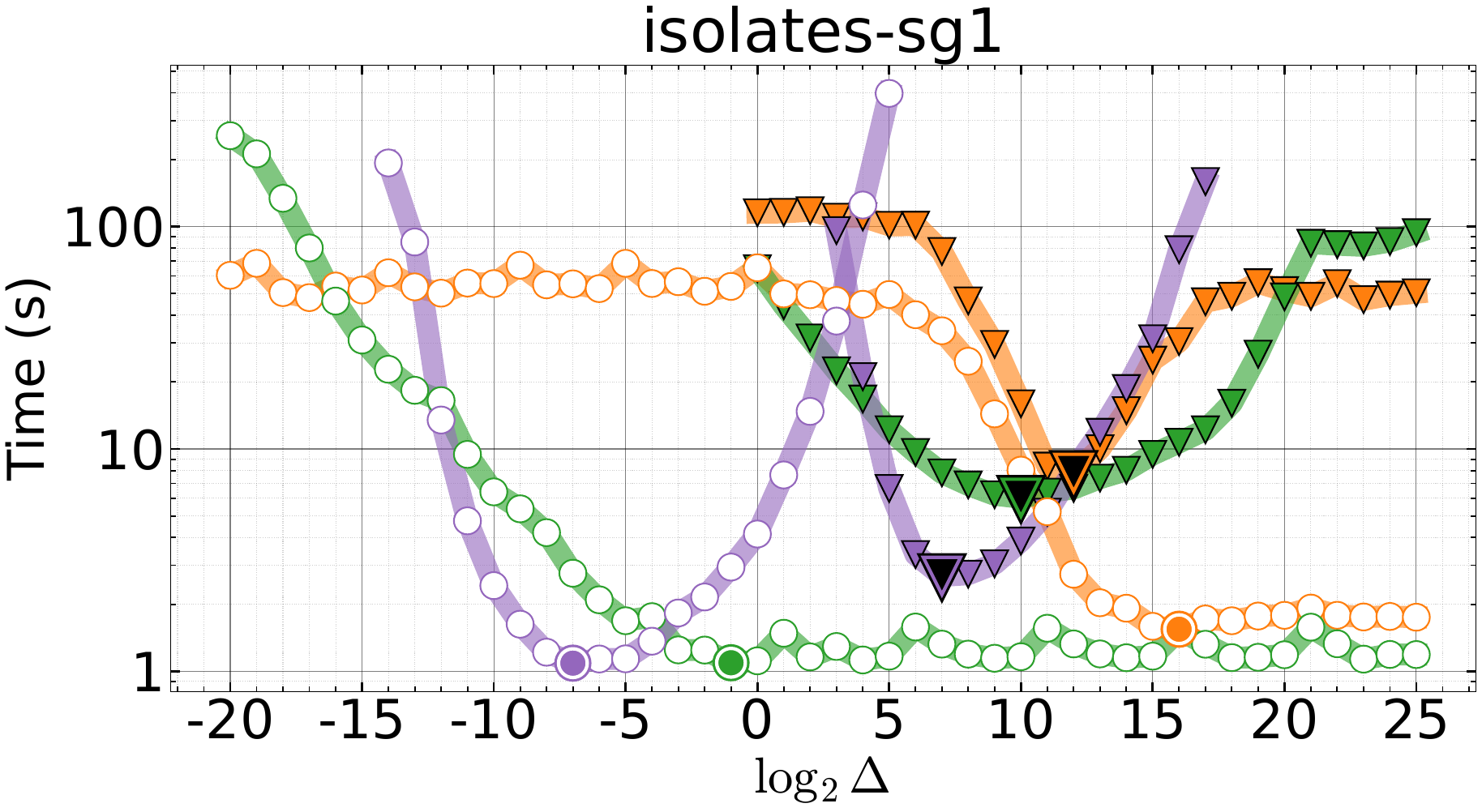}
    \end{subfigure}\hfill
    \begin{subfigure}[b]{0.49\textwidth}
        \centering
        \includegraphics[width=.87\textwidth]{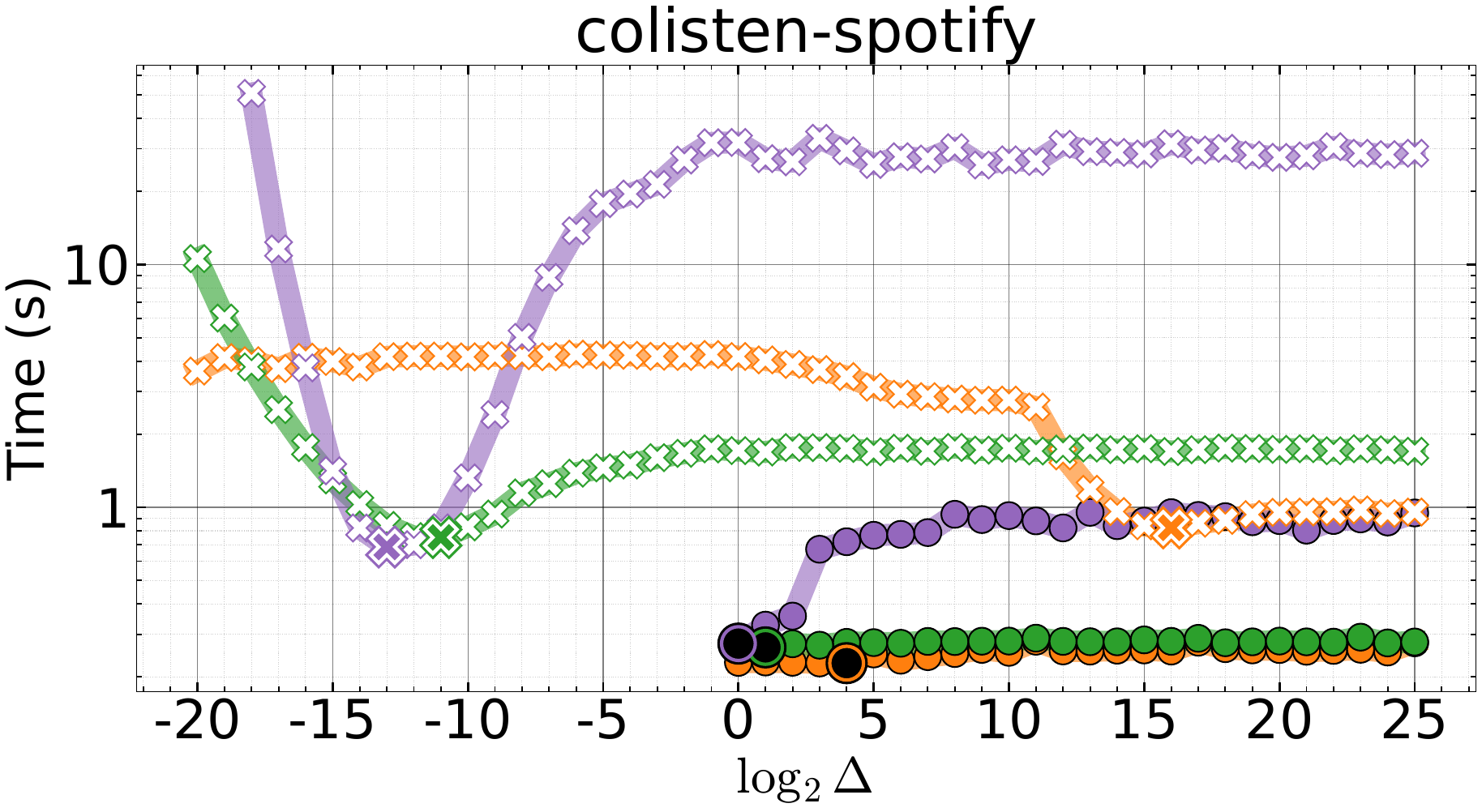}
    \end{subfigure}\hfill
    \begin{subfigure}[b]{0.49\textwidth}
        \centering
        \includegraphics[width=.87\textwidth]{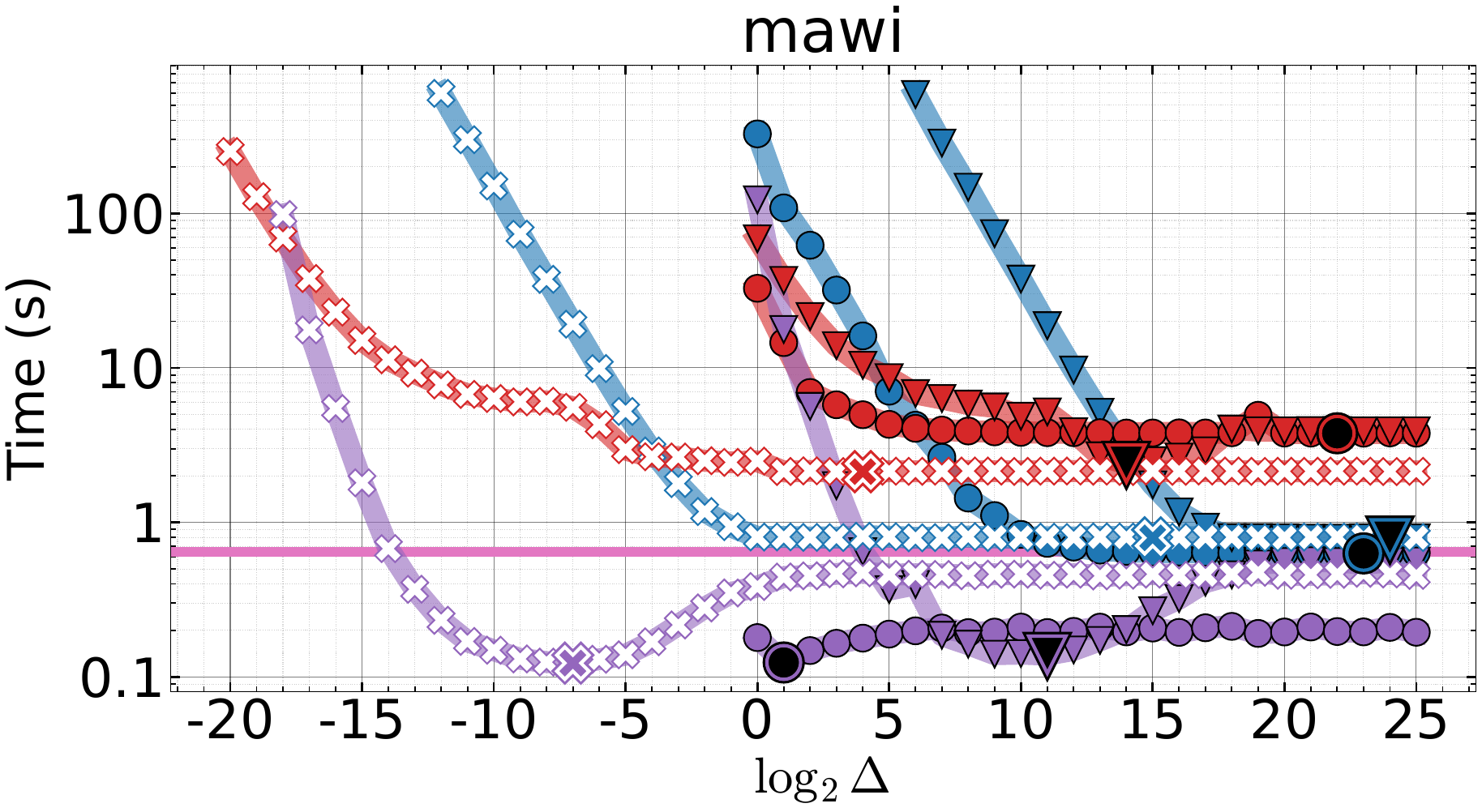}
    \end{subfigure}\hfill
    \caption{Performance across values of $\Delta$ of different parallel SSSP implementations on four representative graphs and weight distributions. The x-axis shows the logarithm of $\Delta$. The best-performing parameter is indicated by a larger and filled marker. Integer weights are indicated with a filled marker, while float weights are indicated with a hollow marker.}
    \label{fig:delta-curves}
    \end{figure*}

To demonstrate the impact of edge weight distributions on the tuning behavior of $\Delta$-based implementations Figure~\ref{fig:delta-curves} shows the performance of the analyzed SSSP algorithms with different weight distributions when sweeping across values of $\Delta$.
Different distributions are indicated by different markers.
By focusing on lines of the same color and comparing the optimal $\Delta$ (indicated by a larger, filled marker), we notice that the optimal $\Delta$ is highly volatile.
Rather than undergoing minor adjustments, this value frequently shifts by several orders of magnitude across weight distributions.
\begin{observation}
Different edge weight distributions dictate distinct optimal values for $\Delta$, often shifting the ideal parameter by several orders of magnitude and completely invalidating the portability of parameters tuned on synthetic graphs.
\end{observation}

As demonstrated in the previous section, no parameterized parallel SSSP implementation is entirely immune to mis-tuning penalties. 
However, \rhostep presents a particularly interesting case. 
The original work introduced the algorithm as a robust alternative to $\Delta$-stepping, suggesting that a consistently large parameter -- such as $\rho=2^{21}$ -- serves as an effective, tuning-free default.
Specifically, the paper notes that configurations between $2^{20}$ and $2^{24}$ almost always remain within a 20\% performance margin of the optimal execution time.
Our experimental analysis reveals a more nuanced picture: the tuning behavior of $\rho$-stepping remains highly coupled to both edge weights and graph topology. 
As highlighted in Figure~\ref{fig:delta-curves}, for natural weights, the \texttt{planet} graph has an optimal $\rho=2^{13}$, which falls far outside the suggested default range.
Similarly, on the \texttt{isolates-sg1} network, even when applying the exact $U_{DONG}$ synthetic distribution evaluated in the original paper, the optimal parameter is found at $\rho=2^{12}$.
In such cases, increasing the value of $\rho$ further incurs notable performance degradation rather than staying within the expected 20\% margin. 
This demonstrates that theoretical parameter portability is ultimately bound by structural and weight-driven constraints.

\begin{observation}
While theoretically robust, the optimal configuration for $\rho$-stepping is still tied to both graph topology and edge weight distribution.
Consequently, relying on a static large default parameter frequently incurs severe performance penalties in practice.
\end{observation}

By looking at the overall curve of the performance with different $\Delta$, we notice that the weight distribution impacts the shape of the curve.
This is visible for both the \texttt{isolates-sg1} and \texttt{colisten-spotify} graphs for the GBBS and \rhostep implementations.
For example, on \texttt{isolates-sg1}, the execution time under the natural weight distribution stabilizes into a plateau for larger $\Delta$ values. 
Conversely, the $U_{DONG}$ distribution creates a clear global minimum for both implementations, after which performance sharply degrades.
This variability directly dictates the complexity and efficiency required of any dynamic $\Delta$-tuning strategy.
For instance, while $\Delta=2^{23}$ has close to optimal performance on GBBS with the natural weight distribution, applying the same $\Delta$ with the $U_{DONG}$ incurs a performance degradation of over an order of magnitude. 
\begin{observation}
The edge weight distribution fundamentally alters the shape of the performance curve across $\Delta$ values, frequently transforming robust execution plateaus into highly sensitive minima that drastically amplify mis-tuning penalties.
\end{observation}

Finally, we highlight an interesting phenomenon occurring within the \texttt{mawi} graph. 
In this specific topology, a single massive hub is connected to 99\% of the other vertices in the graph, which are themselves degree-1 leaves. 
Because work distribution in these frameworks is typically managed at the vertex level rather than the edge level, processing this hub creates severe load imbalance. 
This structural bottleneck dominates the execution, artificially masking the impact of the edge weight distributions. 
This masking effect is clearly visible in the performance of GAP and \dstarstep; across all evaluated weight distributions, these implementations favor an extremely large $\Delta$, inherently regressing their execution model to that of Bellman-Ford. 
This convergence is validated by the parallel Bellman-Ford algorithm, which performs similarly to the optimal configurations of GAP and \dstarstep. 
For clarity in Figure~\ref{fig:delta-curves}, we represent the Bellman-Ford execution time as a single pink horizontal line using only natural weights, as its performance is similar across all weight distributions (as shown in Section~\ref{sec:performance}).
Conversely, Wasp implements a structural optimization that prevents degree-1 leaves from being inserted into the scheduling buckets~\cite{dantonioWaspEfficientAsynchronous2025}, an essential mechanism for achieving high performance on this topology. 
Strikingly, once this high-degree topological challenge is mitigated, the previously hidden effects of the weight distribution re-emerge: Wasp exhibits distinct performance curves and different optimal $\Delta$ values for each individual distribution.

\begin{observation}
Graph topology and edge weights are coupled; topological bottlenecks can force $\Delta$-stepping algorithms to regress to Bellman-Ford execution, masking the impact of the weight distribution.
\end{observation}

\section{Discussion and Conclusion}\label{sec:conclusion}
Parallel SSSP benchmarking has been founded on the assumption that synthetic, uniform weight distributions serve as adequate proxies for real-world data.
Our characterization and performance analysis of seven algorithms denies this assumption, proving that edge weights drive algorithmic performance and parameter tuning.
We have shown how synthetic weight distributions misrepresent real-world SSSP performance, exhibiting large variability of results with artificial speedups and slowdown across classes of graphs.
Furthermore, the weight distributions complicate the tuning of algorithmic parameters, warping the optimal tuning landscape for $\Delta$-based and \rhostep algorithms.

In light of these findings, we distill our analysis into actionable insights for three distinct audiences within the parallel graph processing community:
\paragraph{\textbf{Users}}
For users deploying parallel SSSP algorithms on real-world datasets, our results demonstrate that relying on universally suggested default parameters is a dangerous pitfall. 
Because the tuning landscape is so easily deformed by edge weights, threshold parameters such as $\Delta$ and $\rho$ must be tuned directly on the specific weight distribution of the target graph. 

When selecting an implementation, users must weigh raw performance against tuning effort.
If ease of use and robustness to weight variations are the primary requirements, we suggest utilizing the MultiQueue approach.
If resources allow for meticulous parameter tuning, Wasp delivers the best overall execution time while remaining resilient to diverse weight distributions.
Alternatively, $\Delta^*$-stepping serves as a highly performant secondary option, although with larger sensitivity to weight variations compared to asynchronous algorithms.

\paragraph{\textbf{Performance Analysts}}
For the designers and evaluators of graph systems, our analysis underscores that simple uniform weight distributions are fundamentally not representative of natural weights.
All of the real-world weight distributions used in this study have a log-normal body.
Evaluating algorithms exclusively on synthetic uniform weights frequently misidentifies the most performant algorithm compared to natural weight baselines. 

To bridge the gap between benchmark results and real-world deployment, the community must invest a larger effort into developing sophisticated synthetic weight generation.
We advocate for adopting synthetic distributions that statistically mirror real-world data, such as log-normal distributions or mixed models featuring a log-normal body with a heavy tail.
This, however, requires additional research as the performance of SSSP also depends on the relationship between edge weights and graph topology~\cite{buInterplayTopologyEdge2023}.

\paragraph{\textbf{Algorithm Designers}}
Finally, for algorithm designers, this study highlights that future efforts should focus on decoupling efficiency from parameter choice.
The steep performance degradation observed in statically configured $\rho$-stepping and parallel Bellman-Ford illustrates the fragility of strict synchrony.
Our findings indicate that asynchronous algorithms are more resilient to diverse weight distributions and the highly skewed weights characteristic of real-world networks.
Future designs should prioritize this robustness, pushing toward parameter-free or dynamic thresholding approaches that adapt to high weight skew.

\section*{Acknowledgment}
This work was partially funded by the European Union, Horizon Europe 2021-2027 Framework Programme, Grant No. 101072456, and the UK Research and Innovation, Engineering and Physical Sciences Research Council, Grant No. EP/X029174/1.
The authors used generative AI tools for editing; all ideas,
content, and conclusions are their own.

\bibliographystyle{IEEEtranS}
\bibliography{references}

\clearpage
\appendix
\section{Artifact Appendix}

\subsection{Abstract}

This artifact appendix details the experimental setup, software dependencies, and workflow automation required to reproduce the edge weight characterizations and algorithm performance evaluations presented in the paper. We provide a concise overview of the reproduction workflow, leaving specific execution options to the repository's README files. The artifact enables the reproduction of the primary empirical results reported in the paper, specifically Figures 1--3, Figures 5--6, and Table 3.

\subsection{Artifact check-list (meta-information)}
{\small
\begin{itemize}
  \item {\bf Algorithm: } Parallel Single-Source Shortest Path (SSSP).
  \item {\bf Program: } Wasp, GBBS, MultiQueue, GAPBS, PASGAL ($\Delta^*$-stepping, $\rho$-stepping, parallel Bellman-Ford).
  \item {\bf Compilation: } GCC/Clang with C++17/C++20 and OpenMP support, CMake ($\ge 3.13$), Make, Bazel/Bazelisk.
  \item {\bf Binary: } C++ executables compiled from source via provided setup scripts.
  \item {\bf Data set: } 17 large-scale real-world graphs (Matrix Market, Weighted Edge List, binary GAP format) and 7 synthetic weight distribution schemas.
  \item {\bf Run-time environment: } GNU/Linux x86-64, Python 3.10+ (with virtual environment), \texttt{libnuma}, and \texttt{numactl}.
  \item {\bf Hardware: } Multicore x86\_64 processor (preferably 16+ CPU cores), $\ge 128$~GB system DRAM, $\sim 2$~TB free disk storage.
  \item {\bf Execution: } Orchestrated via Python scripts (\texttt{launch.py}, \texttt{delta-scaling.sh}), supporting local multi-threaded execution and SLURM batch submission.
  \item {\bf Metrics: } Execution time (seconds), speedup, weight sensitivity factor, distribution fitting parameters, and normalized mis-tuning penalty.
  \item {\bf Output: } Execution logs, fitting statistics summary, and graphics reproducing paper figures.
  \item {\bf How much disk space required (approximately)?: } $\sim2$~TB for all graph datasets and binary structures.
  \item {\bf How much time is needed to prepare workflow (approximately)?: } $\sim5$ hours (downloading, decompress, extracting, converting graphs to binary format, and generating synthetic weights).
  \item {\bf How much time is needed to complete experiments (approximately)?: } $\sim 5$--$15$ minutes using pre-computed log files from our experiments; $\sim71$ days for full experimental execution from scratch on a single machine. The number of graphs and implementations evaluated can be reduced to reduce the overall time. 
  \item {\bf Publicly available?: } Yes.
  \item {\bf Code licenses (if publicly available)?: } MIT License.
  \item {\bf Data licenses (if publicly available)?: } Public Domain (\texttt{metaclust}, \texttt{isolates-sg1}); CC-BY 4.0 License (all other datasets).
\item {\bf Archived: \href{https://doi.org/10.5281/zenodo.22046250}{doi.org/10.5281/zenodo.22046251}}

\end{itemize}
}

\subsection{Description}

\subsubsection{How to access}

The artifact source code, configuration scripts, and pre-computed execution logs are available at: \href{https://doi.org/10.5281/zenodo.22046250}{doi.org/10.5281/zenodo.22046251}.

\subsubsection{Hardware dependencies}
A multicore x86-64 processor, 128~GB of DRAM, and 2~TB of available disk space to accommodate raw graph files and binary representations.

\subsubsection{Software dependencies}

A modern 64-bit GNU/Linux distribution (evaluated on Rocky Linux, kernel 4.18.0).
C++ compiler supporting C++17/C++20 and OpenMP.
Build systems: CMake ($\ge 3.13$), GNU Make, and Bazelisk.
Core libraries: \texttt{libnuma} and \texttt{numactl}.
Python 3.10+ with standard virtual environment (\texttt{venv}).
Experiments can optionally run under the SLURM workload manager.

\subsubsection{Data sets}

17 real-world graph datasets spanning web, social, biological, and road networks (\texttt{central-america}, \texttt{australia-oceania}, \texttt{south-america}, \texttt{africa}, \texttt{north-america}, \texttt{asia}, \texttt{europe}, \texttt{planet}, \texttt{archaea}, \texttt{eukarya}, \texttt{isolates-sg1}, \texttt{metaclust}, \texttt{mawi}, \texttt{coauth-aminer}, \texttt{coauth-mag}, \texttt{colisten-spotify}, \texttt{moliere}).
Synthetic weight distributions encompass 6 schemas ($U_{G500}$, $U_{BEAM}$, $U_{DHUL}$, $U_{DONG}$, $N_{PANI}$, $N_{ANON}$).
Dataset URLs and extraction guidelines are provided in the artifact.

\subsection{Installation}

Extract the artifact repository from the provided archive. Following the step-by-step instructions in the README files, configure the required environment variables, setup the Python virtual environment, download and convert the graph datasets to binary format, generate synthetic edge weight distributions and source vertex lists, and compile the C++ SSSP implementations.

\subsection{Experiment workflow}

The experimental workflow is organized into two primary evaluation phases:
\begin{itemize}
    \item {\bf Characterization Analysis:} Computes edge weight frequency distributions across all natural graphs, determines optimal lower bounds $x_{\min}$ for power-law fitting by minimizing Kolmogorov-Smirnov distance. Evaluators should note that finding $x_{\min}$ and validating the power-law hypothesis across large datasets is computationally intensive.
    \item {\bf Performance and Tuning Analysis:} Sweeps stepping parameter values $\Delta$ across combinations of graph datasets, algorithm implementations, and weight distributions. Evaluators may either execute full (or partial) experimental runs from scratch or utilize the provided pre-computed log files from our experiments to reproduce the paper results.
    
\end{itemize}

\subsection{Evaluation and expected results}

The artifact provides parsing and plotting scripts to process execution logs and generate vector graphics reproducing Figures 1--3, Figures 5--6, and Table 3. The empirical characterizations of edge weight distributions will match those reported in Table 3. While absolute execution runtimes may vary depending on the underlying hardware architecture, relative algorithmic speedups, weight sensitivity trends, and optimal $\Delta$ parameter behaviors remain consistent.

\end{document}